\documentclass[twocolumn,english,reprint, prl]{revtex4-1}
\usepackage[T1]{fontenc}
\usepackage[latin9]{inputenc}
\usepackage{geometry}
\usepackage{color}
\usepackage{booktabs}
\usepackage{multirow}
\usepackage{graphicx}
\usepackage{esint}

\makeatletter

\providecommand{\tabularnewline}{\\}

\usepackage{xcolor,colortbl}
\definecolor{Gray}{gray}{0.90}

\makeatother

\usepackage{babel}
\begin{document}

\title{Bumblebee flight in heavy turbulence}

\author{T. Engels $^{1,2}$, D. Kolomenskiy$^{3}$, K. Schneider$^{1}$,
F.-O. Lehmann$^{4}$ and J. Sesterhenn$^{2}$}

\affiliation{$^{1}$ M2P2-CNRS \& Aix-Marseille Université, 38 rue Joliot-Curie,
13451 Marseille cedex 20 France $^{2}$ ISTA, Technische Universität
Berlin, Müller-Breslau-Strasse 12, 10623 Berlin, Germany $^{3}$ Biomechanical
Engineering Laboratory, Chiba University, 1-33 Yayoi-Cho, Inage-Ku,
Chiba-Shi, Chiba, 263-8522 Japan $^{4}$ Universität Rostock, Department
of Animal Physiology, Albert-Einstein-Str. 3, 18059 Rostock, Germany }
\begin{abstract}
High-resolution numerical simulations of a tethered model bumblebee
in forward flight are performed superimposing homogeneous isotropic
turbulent fluctuations to the uniform inflow. Despite tremendous variation
in turbulence intensity, between 17\% and 99\% with respect to the
mean flow, we do not find significant changes in cycle-averaged aerodynamic
forces, moments or flight power when averaged over realizations, compared
to laminar inflow conditions. The variance of aerodynamic measures,
however, significantly increases with increasing turbulence intensity,
which may explain flight instabilities observed in freely flying bees. 
\end{abstract}
\maketitle
Insect flight currently receives considerable attention from both
biologists and engineers. This growing interest is fostered by the
recent trend in miniaturization of unmanned air vehicles that naturally
incites reconsidering flapping flight as a bio-inspired alternative
to fixed-wing and rotary flight. For all small flyers it is challenging
to fly outdoors in an unsteady environment, and it is essential to
know how insects face that challenge. 
\begin{figure*}
\begin{centering}
\includegraphics[width=1\textwidth]{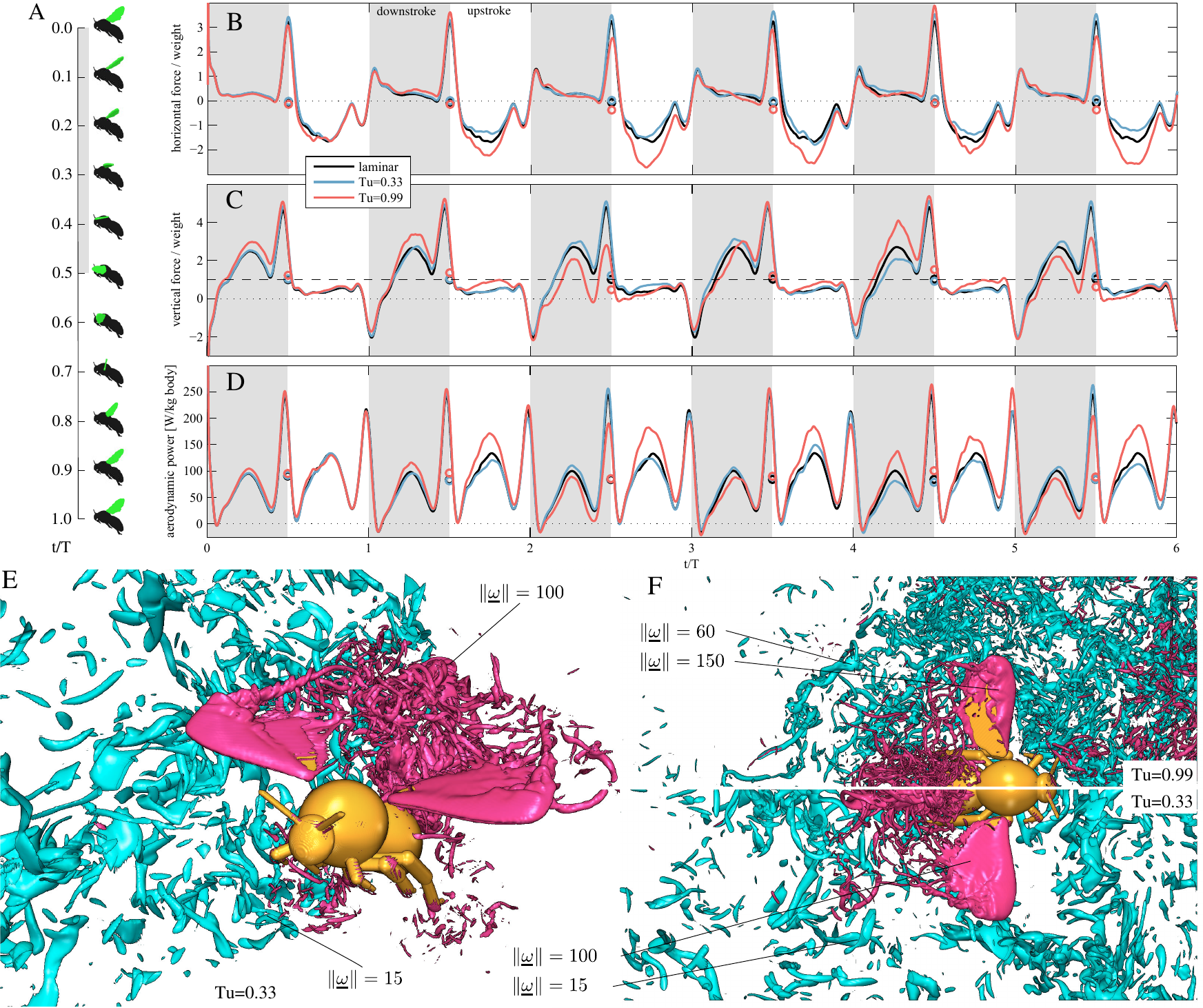}
\par\end{centering}

\caption{Bumblebee in turbulent flow. (A) Visualization of the prescribed wingbeat,
where $T$ is period time. (B-D) Time evolution of horizontal (B)
and vertical (C) force, and aerodynamic power (D) under laminar, moderately
turbulent ($Tu=0.33$) and highly turbulent ($Tu=0.99$) conditions.
Circular markers represent cycle-averaged values. (E-F) Flow visualization
by means of isosurfaces of normalized vorticity magnitude $\left\Vert \underline{\omega}\right\Vert $.
(E) Perspective view for a realization with $Tu=0.33$. The purple
and blue isosurfaces visualize stronger and weaker vortices, respectively,
and weaker vortices are shown only for $3.7R\leq y\leq4R$. (F) Top
view, with the upper half showing flow at elevated turbulent ($Tu=0.99$),
and the lower half at moderate turbulent ($Tu=0.33$) intensity. Weaker
vortices, i.e., smaller values of $\left\Vert \underline{\omega}\right\Vert $,
are shown only for $0\leq z\leq0.3R$.\label{fig:fancy-fig}}
\end{figure*}

Field studies show variations of insect behavior with changing weather
conditions, including the atmospheric turbulence \cite{Swartz2008}.
Earlier laboratory research on aerodynamics of insect flight assumed
quiescent air, and only some more recent experiments focused on the
effect of different kinds of unsteady flows. The behavior of orchid
bees flying freely in a turbulent air jet has been studied in \cite{Combes2009}.
The authors found that turbulent flow conditions have a destabilizing
effect on the body, most severe about the animal's roll axis. In
response to this flow, bees try to compensate the induced moments
by an extension of their hindlegs, increasing the roll moment of inertia.
Interaction of bumblebees with wake turbulence has also been considered
in \cite{Ravi2013}. These experiments were performed in a von Kármán-type
wake behind cylinders. The bees displayed large rolling motions, pronounced
lateral accelerations, and a reduction in their upstream flight speed.
In \cite{Vance2013} a comparative study on the sensitivity of honeybees
and stalk-eye flies to localized wind gusts was performed. The study
found that bees and stalk-eye flies respond differently to aerial
perturbations, either causing roll instabilities in bees or significant
yaw rotations in stalk-eye flies. In \cite{Ortega-Jimenez2013} feeding
flights of hawkmoths in vortex streets past vertical cylinders were
analyzed. Depending on distance of the animal from the cylinder and
cylinder size, destabilizing effects on yaw and roll and a reduction
in the animal's maximum flight speed have been observed. Kinematic
responses to large helical coherent structures were also found in
hawkmoths flying in a vortex chamber \cite{Ortega-Jimenez2014}. A
study on the energetic significance of kinematic changes in hummingbird
feeding flights further demonstrated a substantial increase in metabolic
rate during flight in turbulent flows, compared to flight in undisturbed
laminar inflow \cite{Ortega-Jimenez2014a,Ravi2015}. All studies reported
significant changes in the behavior of insects when they fly in turbulent
flows and incite the question if, and how, the efficiency of flapping
wings changes. It is critical to understand whether the aerodynamic
challenge insects face when flying through turbulence is due to the
elevated power requirements and reduced force production, or rather
limited capacity of flight controls. Experiments with freely flying
animals involve complex, sensory-dependent changes in wing kinematics
and wing-wake interaction. To isolate specific effects of turbulence
on aerodynamic mechanisms and power expenditures in flight, direct
numerical simulations are well suited tools. However, to determine
statistical moments of the forces and torques acting on the insect,
a sufficient number of flow realizations needs to be computed, owing
to the generic randomness of turbulence. 

In this Letter, we present the first direct numerical simulations
of insect flight in fully developed turbulence, using a model bumblebee
based on \cite{Dudley1990}. We address the question of how turbulence
alters forces, moments and power expenditures in flapping flight.
Since bumblebees are all-weather foragers, they encounter a particularly
large variety of natural flow conditions \cite{Ravi2013}. We consider
our model bumblebee in forward flight at 2.5 m/s, flapping its rigid
wings (Fig. \ref{fig:fancy-fig}A) at a Reynolds number of 2042. To
conduct the simulation, we designed a `numerical wind tunnel' and
placed the animal in a $6R\times4R\times4R$ large, virtual, rectangular
box, where $R=13.2$ mm is the wing length. The computational domain
is discretized with 680 million grid points and the incompressible
three-dimensional Navier--Stokes equations are solved by direct numerical
simulation \cite{Engels2015a}. An imposed mean inflow velocity accounts
for the forward flight speed of the tethered insect, with superimposed
velocity fluctuations in the turbulent cases. Since the actual properties
of these aerial perturbations depend on a large number of parameters,
we model them by homogeneous isotropic turbulence (HIT) \cite{Rogallo1981,Kaneda2003}.
This is a reasonable assumption for the small turbulent scales relevant
to insects. In addition, HIT is a well established type of turbulence
which reduces the set of parameters to the turbulent Reynolds number
$R_{\lambda}$. Insect flight can thus be studied from laminar to
fully-developped turbulent flow conditions, yielding time series of
aerodynamic measures (Fig. \ref{fig:fancy-fig}B-D), as well as the
flow data (Fig. \ref{fig:fancy-fig}E-F). Further details on the model
and the simulations can be found in the supplementary material \cite{supplmat}.

\begin{figure}
\begin{centering}
\includegraphics[width=1\columnwidth]{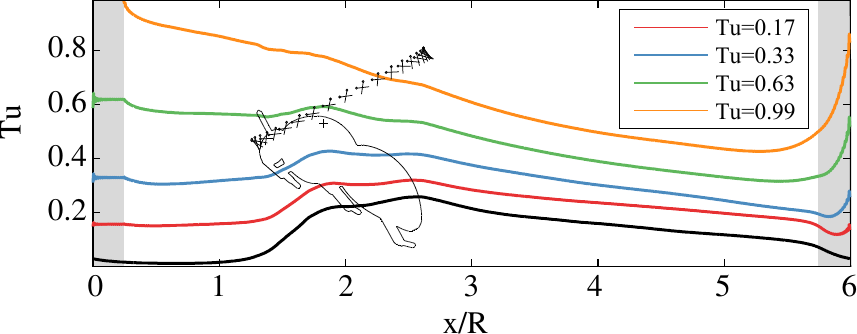}
\par\end{centering}

\caption{Slab averaged turbulence intensity as a\textcolor{red}{{} }function
of the axial coordinate, with the insect drawn to scale for orientation.
Black line is the laminar case. The gray shaded areas mark regions
where the in- and outflow is imposed.\label{fig:slab-avg-turbulence-intensity}}
\end{figure}

First, we focus on the wake pattern generated by the insect in laminar
inflow. This case serves as reference for the turbulence simulations
and provides quantitative data on vortical flow generated by the flapping
wings. Fig. \ref{fig:fancy-fig}B-D shows how body weight-normalized
lift and thrust, and body mass specific aerodynamic power vary throughout
the flapping cycles. Force and power peak during the stroke reversals,
as observed in \cite{Dickinson1999}. The cycle-averaged flight forces
obtained from this simulation are summarized in table \ref{tab:Data-from-the-experiments}.
The data show that the bumblebee model produces lift that matches
the weight to within 2\%, but 8\% more thrust than required to compensate
for free stream velocity. These slight discrepancies result from the
uncertainty of the input parameters. The aerodynamic power required
to actuate the wings is 84 W/kg body mass. This is larger than the
value reported in \cite{Dudley1990a} (56 W/kg body mass), which may
be explained by the differences in the wing kinematics and the aerodynamic
models employed. Mean moments about the three rotational body axes
do not significantly differ from zero suggesting a torque balanced
force production. The turbulence intensity, $Tu=u'/u_{\infty}$, is
the root mean square (RMS) of velocity fluctuations normalized to
flight velocity. Fig. \ref{fig:slab-avg-turbulence-intensity} presents
slab-averaged turbulence intensity $\left\langle Tu\right\rangle =\int_{y_{0}-1.3R}^{y_{0}+1.3R}\int_{z_{0}-1.3R}^{z_{0}+1.3R}Tu(x,y,z)\mathrm{d}y\mathrm{d}z/\left(2.6R\right)^{2}$
as a function of the downstream distance. The black line corresponds
to the laminar case. The subdomain used for averaging is centered
around the insect ($y_{0}=2R,$ $z_{0}=2R$). Complementary 3D visualizations
of the wake can be found in the supplementary material \cite{supplmat}.
The data show that the bumblebee model generates relative intensities
of 25\% at the wings and approximately 16\% at five wing lengths downstream
distance. This finding indicates that relevant turbulence intensities
are larger than 16\%, which is well above the inflow turbulence considered
in investigations concerning airfoils, typically below 1\% \cite{Mueller1983}.

\begin{figure}
\begin{centering}
\includegraphics[width=1\columnwidth]{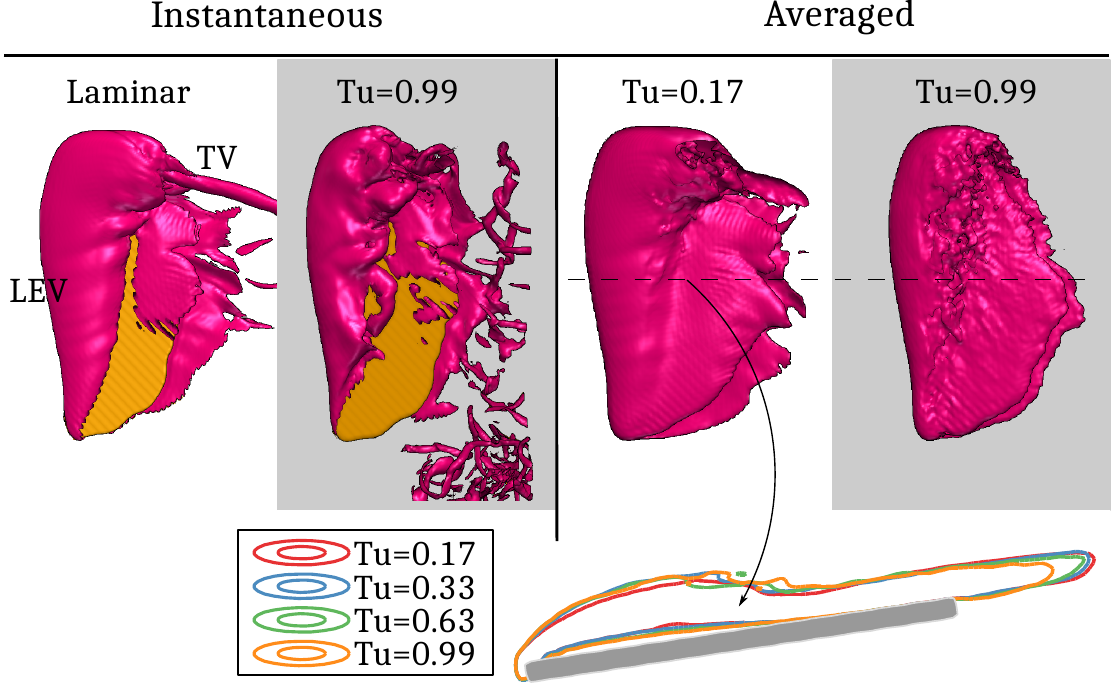}
\par\end{centering}

\caption{\label{fig:Phase--and-ensemble-LEV}Top: Isosurface of normalized
absolute vorticity, $\left\Vert \underline{\omega}\right\Vert =100$,
in the vicinity of the right wing at $t/T=0.3$. Snapshots of instantaneous
vorticity distribution during laminar and turbulent inflow is shown
on the left. Phase- and ensemble averaged vorticity from $16$ and
$108$ wing beats and at $Tu=0.17$ and $Tu=0.99$, respectively,
is shown on the right. Bottom: averaged $\left\Vert \underline{\omega}\right\Vert =50$
isolines at mid-span for all values of $Tu$. The leading edge vortex
persists on average even under strongest inflow perturbations.}
\end{figure}

Second, we study the model insect in turbulent inflow considering
four different turbulence intensities, with turbulent Reynolds numbers
$R_{\lambda}=\lambda u'/\nu$ ranging from $90$ to $228$. Here $u'$
is the RMS velocity, $\lambda$ the Taylor micro scale, varying between
$0.25R$ and $0.1R$, and $\nu$ is the kinematic viscosity of air.
The properties of the inflow data are summarized in table \ref{tab:Parameters-of-numerical-experiments}.
For all tested intensities, the Kolmogorov length scale of small,
dissipating eddies, $\ell_{\eta}$, is significantly smaller than
the wing length. The length scale of energy carrying structures, the
integral scale $\Lambda$, is similar to the wing length. The latter
is expected to maximize the impact of turbulence on the insect, while
the former suggests that all vortices generated by the insect interact
nonlinearly with inflow perturbations. To obtain statistically reliable
mean values and variances, we perform $N_{R}$ simulations. Fig. \ref{fig:fancy-fig}E-F
illustrates the flow under turbulent inflow conditions for $Tu=0.33$
and $0.99$ relative intensity. It shows that weak turbulence is associated
with relatively coarse flow structures in the inflow. In contrast,
flow patterns near the wings are similar in size and intensity to
the structures present in the inflow at strong turbulence. The streamwise
slab-averaged turbulence intensity (Fig. \ref{fig:slab-avg-turbulence-intensity})
is increased by the flapping wings for $Tu$ equal to $0.17$ and
$0.33$, while it remains constant or is decreased for $Tu$ equal
to $0.63$ and $0.99$. The lower two values can thus be referred
to as mild turbulence.

The considered range of turbulent Reynolds numbers covers the flow
regime that a bee typically encounters in its natural habitats. Bumblebees
have been reported to fly at wind speeds of 8 m/s \cite{Wolf1999}.
At this speed, habitats with cylindrical trees of about 10 cm in diameter
yield turbulent Reynolds numbers in the range considered here. Fig.
\ref{fig:fancy-fig}B-D shows that lift, thrust and power of single
simulation runs at turbulent conditions differ from the measures obtained
for the laminar case. However, the generic features of the data, i.e.,
the location of peaks and valleys are similar under all tested flow
conditions, see section IA of supporting material. Wingbeat-averaged
and ensemble-averaged data including statistics are shown in table
\ref{tab:Data-from-the-experiments}. The mean values demonstrate
only negligible differences between turbulent and laminar flow conditions
and even at the strongest turbulent perturbation, the bumblebee model
generates mean aerodynamic forces close to those derived in unperturbed
inflow, at virtually the same energetic cost. This aerodynamic robustness
of insect wings is in striking contrast to the properties of streamlined
airfoils that are highly sensitive to the laminar-turbulent transition
\cite{Mueller1983}. Fig. \ref{fig:Phase--and-ensemble-LEV} shows
the vortical structure at the wing at $t/T=0.3$, represented by the
$\left\Vert \underline{\omega}\right\Vert =100$ isosurface of normalized
vorticity $\underline{\omega}=\left(\nabla\times\underline{u}\right)/f$.
The laminar case is a snapshot of the flow field, while turbulent
data are phase-averaged over $N_{w}$ independent strokes for each
value of $Tu$ (see table \ref{tab:Parameters-of-numerical-experiments}).
Although turbulence alters shape and size of the wing's tip vortex,
the leading edge vortex remains visible in phase-averaged flow fields
even at maximum inflow turbulence intensity.

\begin{table*}
\begin{centering}
{\small{}}%
\begin{tabular*}{1\textwidth}{@{\extracolsep{\fill}}ccccccc}
\hline 
\multirow{1}{*}{{\small{}\cellcolor{Gray}$Tu$}} & \multicolumn{1}{c}{{\large{}\strut }{\small{}}%
\begin{minipage}[t]{2cm}%
{\small{}Forward force $F_{h}$}%
\end{minipage}} & \multicolumn{1}{c}{{\small{}\cellcolor{Gray}}%
\begin{minipage}[t]{2cm}%
{\small{}Vertical force $F_{v}$}%
\end{minipage}} & \multicolumn{1}{c}{{\small{}}%
\begin{minipage}[t]{2cm}%
{\small{}Aerodynamic power $P_{\mathrm{aero}}$}%
\end{minipage}} & \multicolumn{1}{c}{{\small{}\cellcolor{Gray}}%
\begin{minipage}[t]{2cm}%
{\small{}Moment $M_{x}$ (roll)}%
\end{minipage}} & \multicolumn{1}{c}{{\small{}}%
\begin{minipage}[t]{2cm}%
{\small{}Moment $M_{y}$ (pitch)}%
\end{minipage}} & \multicolumn{1}{c}{{\small{}\cellcolor{Gray}}%
\begin{minipage}[t]{2cm}%
{\small{}Moment $M_{z}$ (yaw)}%
\end{minipage}}\tabularnewline
\hline 
{\small{}\cellcolor{Gray}$0$} & {\large{}\strut }{\small{}$-0.08^{\pm0.0}\pm0.0$} & {\small{}\cellcolor{Gray}$1.02^{\pm0.0}\pm0.0$} & {\small{}$84.05^{\pm0.0}\pm0.0$} & {\small{}\cellcolor{Gray}$0.00^{\pm0.0}\pm0.0$} & {\small{}$0.01^{\pm0.0}\pm0.0$} & {\small{}\cellcolor{Gray}$0.00^{\pm0.0}\pm0.0$}\tabularnewline
\hline 
{\small{}\cellcolor{Gray}$0.17$} & {\large{}\strut }{\small{}$-0.10^{\pm0.04}\pm0.08$} & {\small{}\cellcolor{Gray}$1.04^{\pm0.09}\pm0.18$} & {\small{}$83.72^{\pm1.77}\pm3.61$} & {\small{}\cellcolor{Gray}$-0.01^{\pm0.01}\pm0.03$} & {\small{}$+0.00^{\pm0.02}\pm0.03$} & {\small{}\cellcolor{Gray}$-0.01^{\pm0.02}\pm0.03$}\tabularnewline
\hline 
{\small{}\cellcolor{Gray}$0.33$} & {\large{}\strut }{\small{}$-0.06^{\pm0.09}\pm0.18$} & {\small{}\cellcolor{Gray}$1.10^{\pm0.10}\pm0.21$} & {\small{}$85.02^{\pm2.03}\pm4.14$} & {\small{}\cellcolor{Gray}$-0.01^{\pm0.04}\pm0.08$} & {\small{}$-0.01^{\pm0.03}\pm0.06$} & {\small{}\cellcolor{Gray}$+0.04^{\pm0.02}\pm0.05$}\tabularnewline
\hline 
{\small{}\cellcolor{Gray}$0.63$} & {\large{}\strut }{\small{}$+0.02^{\pm0.10}\pm0.29$} & {\small{}\cellcolor{Gray}$1.04^{\pm0.13}\pm0.40$} & {\small{}$83.32^{\pm3.13}\pm9.57$} & {\small{}\cellcolor{Gray}$-0.02^{\pm0.04}\pm0.12$} & {\small{}$+0.02^{\pm0.04}\pm0.12$} & {\small{}\cellcolor{Gray}$+0.07^{\pm0.04}\pm0.13$}\tabularnewline
\hline 
{\small{}\cellcolor{Gray}$0.99$} & {\large{}\strut }{\small{}$-0.10^{\pm0.07}\pm0.37$} & {\small{}\cellcolor{Gray}$1.01^{\pm0.10}\pm0.54$} & {\small{}$85.44^{\pm1.98}\pm10.47$} & {\small{}\cellcolor{Gray}$+0.01^{\pm0.04}\pm0.19$} & {\small{}$-0.04^{\pm0.03}\pm0.13$} & {\small{}\cellcolor{Gray}$-0.03^{\pm0.04}\pm0.21$}\tabularnewline
\hline 
\end{tabular*}
\par\end{centering}{\small \par}

\caption{Aerodynamic forces, power and moments obtained in the numerical experiments.
Forces are normalized by the weight $mg$, moments by $mgR$, power
is given in W/kg body mass. Values are given by mean value $\overline{x}$,
95\% confidence interval $\delta_{95}$ and standard deviation $\sigma$
in the form $\overline{x}^{\pm\delta_{95}}\pm\sigma$.\label{tab:Data-from-the-experiments}}
\end{table*}

Previous studies highlighted that turbulent flows may destabilize
body posture of an insect \cite{Combes2009}. Roll, in particular,
is prone to instability because the roll moment of inertia is approximately
four times smaller than about the other axes. Our results in table
\ref{tab:Data-from-the-experiments} show that mean aerodynamic moments
about yaw, pitch, and roll axes do not change with increasing turbulence.
However, we observe characteristic changes in moment fluctuation.
Assuming that during perturbation the insect begins to rotate from
rest at time $t_{0}$, we may approximate the final angular roll velocity
from 
\begin{equation}
\Omega_{roll}(t_{0}+\tau)=\frac{1}{I_{roll}}\int_{t_{0}}^{t_{0}+\tau}M_{roll}(t)\,\mathrm{d}t,\label{eq:roll}
\end{equation}
with $M_{roll}$ the roll moment, $I_{roll}$ the roll moment of inertia
with respect to the body $x$-axis, and $\tau$ the response delay
(see below) \cite{Fry2003}. The maximum turbulence-induced roll velocity
that a freely flying bumblebee encounters depends on the reaction
time of the animal in response to changes in body posture. Many insects
compensate for posture perturbations by asymmetrically changing their
wing stroke. Previous studies on freely flying honeybees reported
response delays of approximately 20 ms or 4.5 stroke cycles, suggesting
the use of ocellar pathways for body stability reflexes in this species
\cite{Vance2013}.

\begin{figure}
\begin{centering}
\includegraphics[width=1\columnwidth]{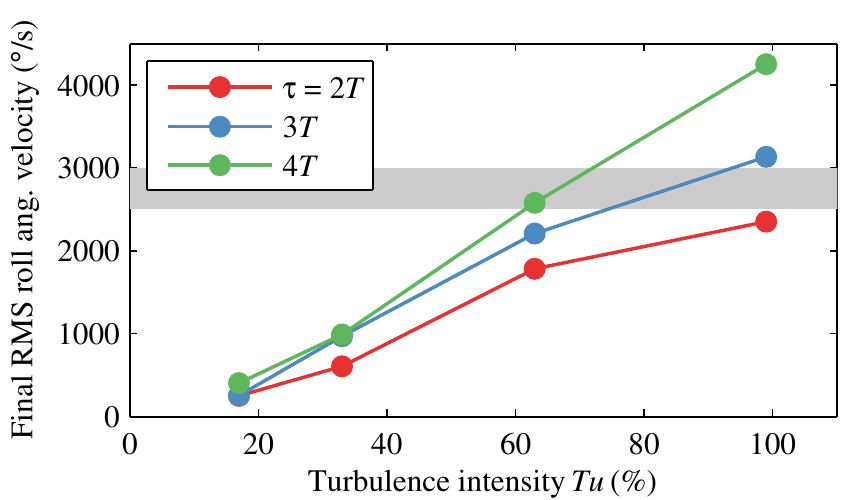} 
\par\end{centering}

\caption{\label{fig:ang_velo}RMS value of the final roll angular velocity
$\Omega_{roll}$ versus the inflow turbulence intensity, calculated
over all flow realizations. Colors correspond to different response
delay times $\tau$. The gray shaded area represents the limit of
sensor saturation, estimated from the behavioral measurements available
for honeybees \cite{Vance2013} and fruitfies \cite{Ristroph2010}.}
\end{figure}

To predict the maximum delay that allows a bumblebee to recover from
turbulence-induced roll, the response delay in equation (\ref{eq:roll})
is set to $\tau=2$, $3$, and $4$ stroke periods. Fig. \ref{fig:ang_velo}
shows how the RMS final roll velocity increases under these conditions
with increasing turbulence intensity. Previous behavioral measurements
provide an estimate of the body angular velocity from which insects
can restabilize in free flight. Fig. \ref{fig:ang_velo} thus predicts
that bumblebees recover form turbulence-induced roll motions up to
$Tu=0.63$ assuming response delays between two and four stroke periods.
In contrast, posture recovery at $Tu=0.99$ requires reduced reaction
times of not more than two cycle periods, implying that bumblebees
cannot achieve stable flight at $Tu=0.99$. This conclusion is consistent
with experimental observations of orchid bees crashing in strongly
turbulent flows when flying freely \cite{Combes2009}.

%41 (caption) + 316 words
% = 357 words

\begin{table}
\begin{centering}
\begin{tabular*}{1\columnwidth}{@{\extracolsep{\fill}}@{\extracolsep{\fill}}ccccccc}
\toprule 
$R_{\lambda}$  & $Tu$  & $\ell_{\eta}$  & $\lambda$  & $\Lambda$  & $N_{R}$  & $N_{w}$\tabularnewline
\midrule
\midrule 
$90.5$  & $0.17$  & $0.013$  & $0.246$  & $0.772$  & 4  & 16\tabularnewline
\midrule 
$130.1$  & $0.33$  & $0.008$  & $0.179$  & $0.782$  & 4  & 16\tabularnewline
\midrule 
$177.7$  & $0.63$  & $0.005$  & $0.129$  & $0.759$  & 9  & 36\tabularnewline
\midrule 
$227.9$  & $0.99$  & $0.004$  & $0.105$  & $0.759$  & 27  & 108\tabularnewline
\bottomrule
\end{tabular*}
\par\end{centering}

\caption{Parameters of inflow turbulence used in simulations. The Kolmogorov
length scale $\ell_{\eta}$, the Taylor micro $\lambda$ and the integral
scale $\Lambda$ are normalized by the wing length $R$. For each
value of $Tu$, a number of $N_{R}$ realizations has been performed,
yielding in total $N_{w}$ statistically independent wingbeats.\label{tab:Parameters-of-numerical-experiments}}
\end{table}

In conclusion, high-resolution numerical experiments of a bumblebee
in perturbed forward flight highlighted several unexpected results
with respect to alterations in aerodynamic forces, flight stability,
and aerodynamic power expenditures. The simulations imply that even
strongest background turbulence does not vitally harm structure and
efficacy of the lift-enhancing leading edge vortex and thus averaged
forces and moments are almost identical compared to laminar flow conditions.
Turbulent inflow conditions are thus of little significance for the
overall flight performance of an animal in tethered flight. However,
these fluctuations cause temporal transient effects. Thus, in a freely
flying insect in which the body may rotate, absolute angular velocities
about yaw, pitch, and roll axes might reach elevated values, which
in turn would require decreasing reaction response delays for body
stabilization with increasing turbulence. Owing to its small moment
of inertia, roll is especially prone to turbulence-induced fluctuations. 

An important consequence of body roll is the deflection of the wingbeat-averaged
resultant aerodynamic force from the vertical direction. Thus, at
large roll angles, the animal must increase the magnitude of this
force so that its vertical component can support the weight of the
insect, at the cost of larger aerodynamic power. Incidentally, it
has been reported that hummingbirds increase the wingbeat frequency
and amplitude \cite{Ravi2015}. The finding that an increase in inflow
turbulence intensity has no significant effect on power expenditures
for tethered flight is surprising and significant with respect to
flight endurance and migration of insects. Since it has been suggested
that flight of insects is limited by power rather than force production
\cite{Ellington1999}, any biological and physical mechanisms that
help an insect to limit its wing and body drag-dependent power expenditures
is of great value and may increase the animal's biological fitness.

As the leading edge vortex is a common feature in many flapping flyers,
we expect these conclusions to generally hold also at different flight
speeds and for other species as well. This is also suggested by our
supplementary results \cite{supplmat} obtained with different morphology,
kinematics and Reynolds number.

We thank H. Liu and M. Maeda for their advice on the kinematic modeling
of flapping wings, and S. Ravi for many valuable comments on the manuscript.
This work was granted access to the HPC resources of Aix-Marseille
Université financed by the project Equip@Meso (ANR-10-EQPX-29-01)
and IDRIS under project number i20152a1664. TE, KS and JS acknowledge
financial support from DFH-UFA, and DK from the JSPS postdoctoral
fellowship P15061.

\bibliographystyle{apsrev4-1}
\nocite{*}
\bibliography{refs}

\end{document}

\title{Bumblebee flight in heavy turbulence: Supplementary material}

\maketitle

\section{Supporting Results}

\subsection{Detailed analysis of flight measures}

In the following we provide the time evolution of the lift force,
aerodynamic power and pitch moment, for each independent wingbeat
in the turbulent cases with $Tu=0.17...0.99$ ($N_{w}=16...108$ ).
For comparison, the corresponding evolution in the laminar case is
also shown, as well as the average over all turbulent realizations
for a given $Tu$. The latter two curves overlap significantly, illustrating
that the ensemble-averaged wing force/moment/power is almost the same
as in the laminar case for all times $t/T$, and not only its integral
value.

\begin{figure}
\begin{centering}
\includegraphics[width=0.5\textwidth]{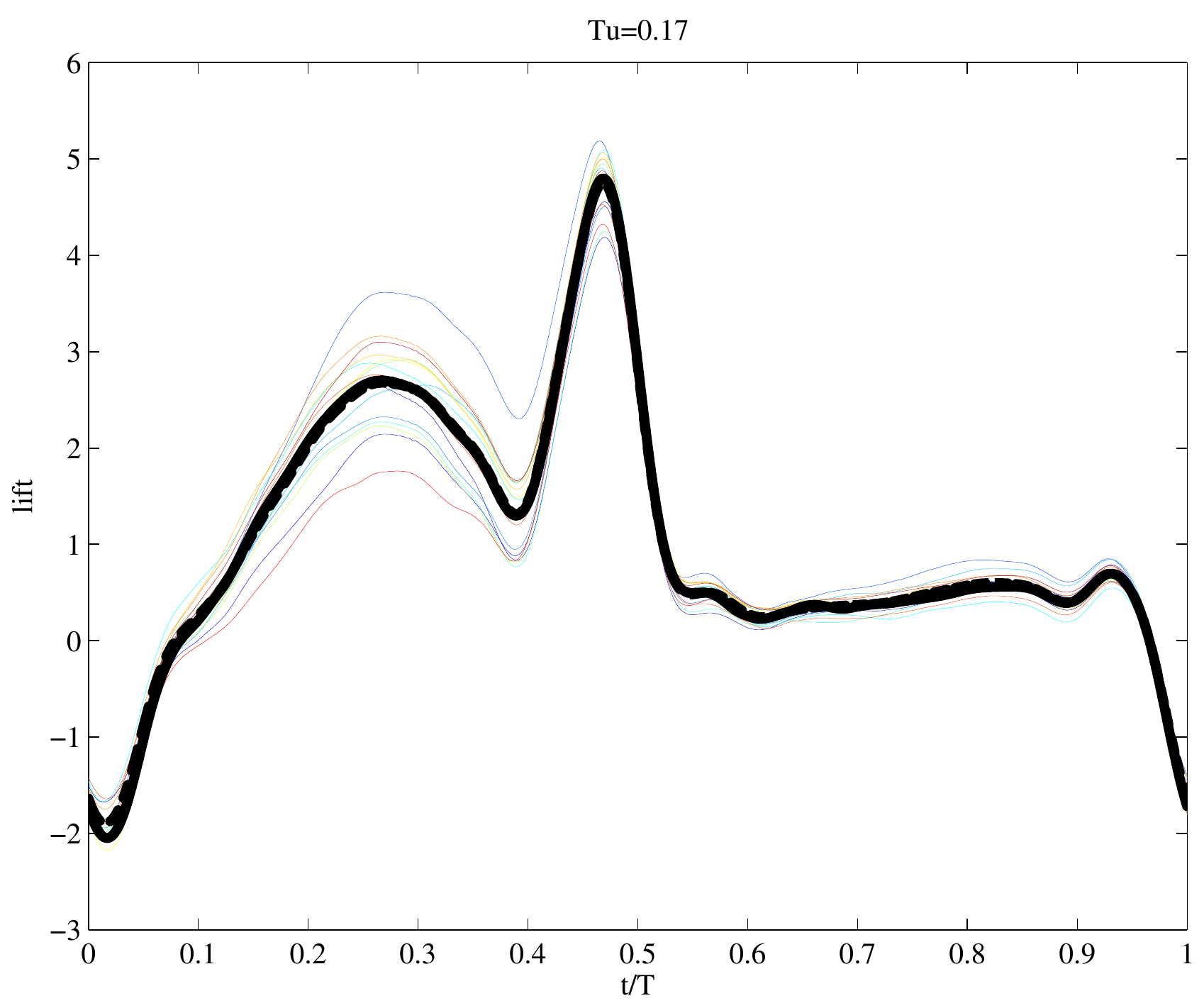}\includegraphics[width=0.5\textwidth]{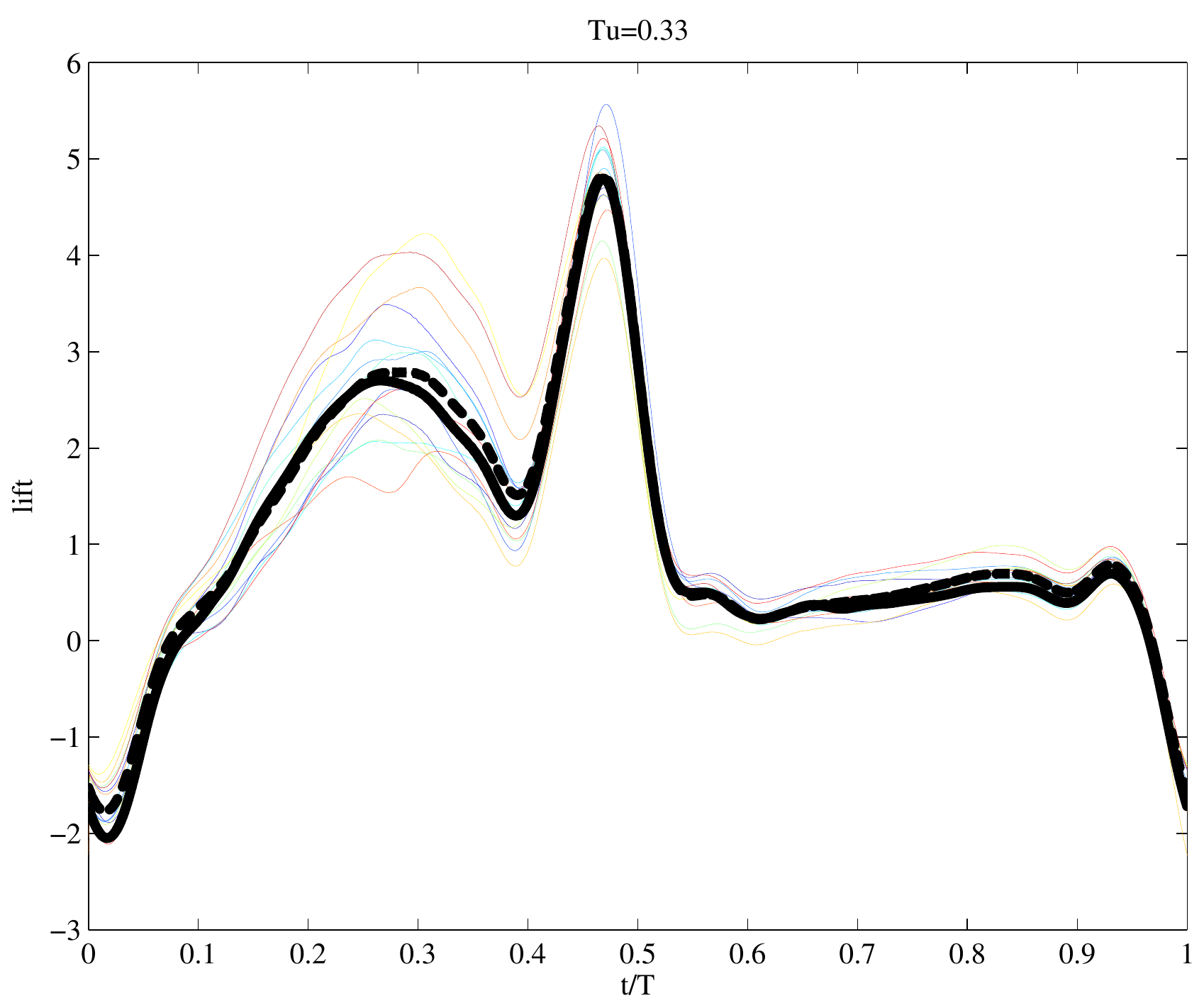}
\par\end{centering}

\begin{centering}
\includegraphics[width=0.5\textwidth]{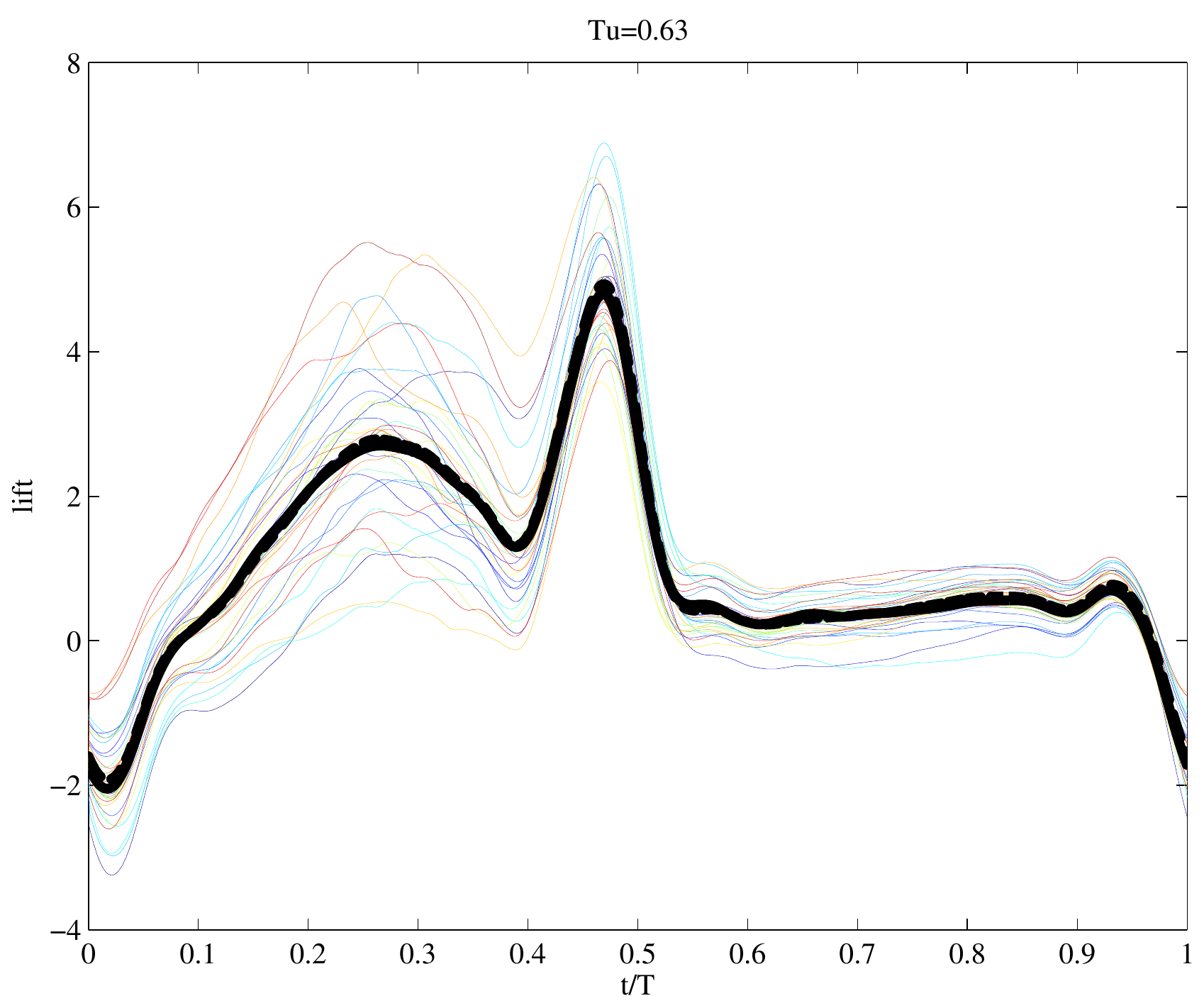}\includegraphics[width=0.5\textwidth]{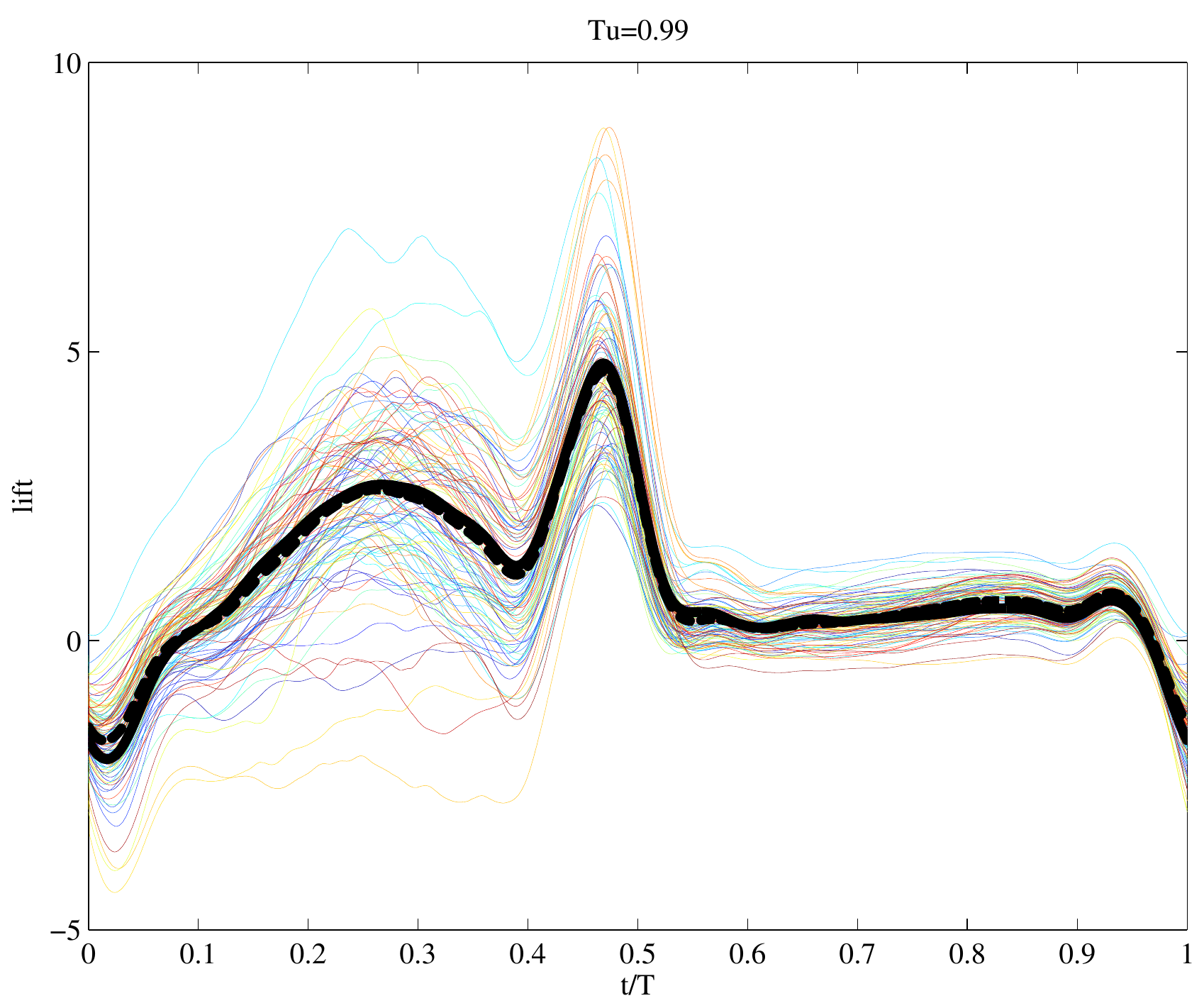}
\par\end{centering}

\caption{Lift force (normalized by weight force) as a function of wing beat
fraction time $t/T$, for different values of $Tu=\left\{ 0.17,0.33,0.63,0.99\right\} $.
Thin colored lines represent individual realizations for the given
turbulence intensity. The average over all turbulent realizations
for the given $Tu$ is represented by the thick black dashed line
and the laminar case by the thick black continuous line. The averaged
turbulent and laminar curves virtually coincide.}
\end{figure}

\begin{figure}
\begin{centering}
\includegraphics[width=0.5\textwidth]{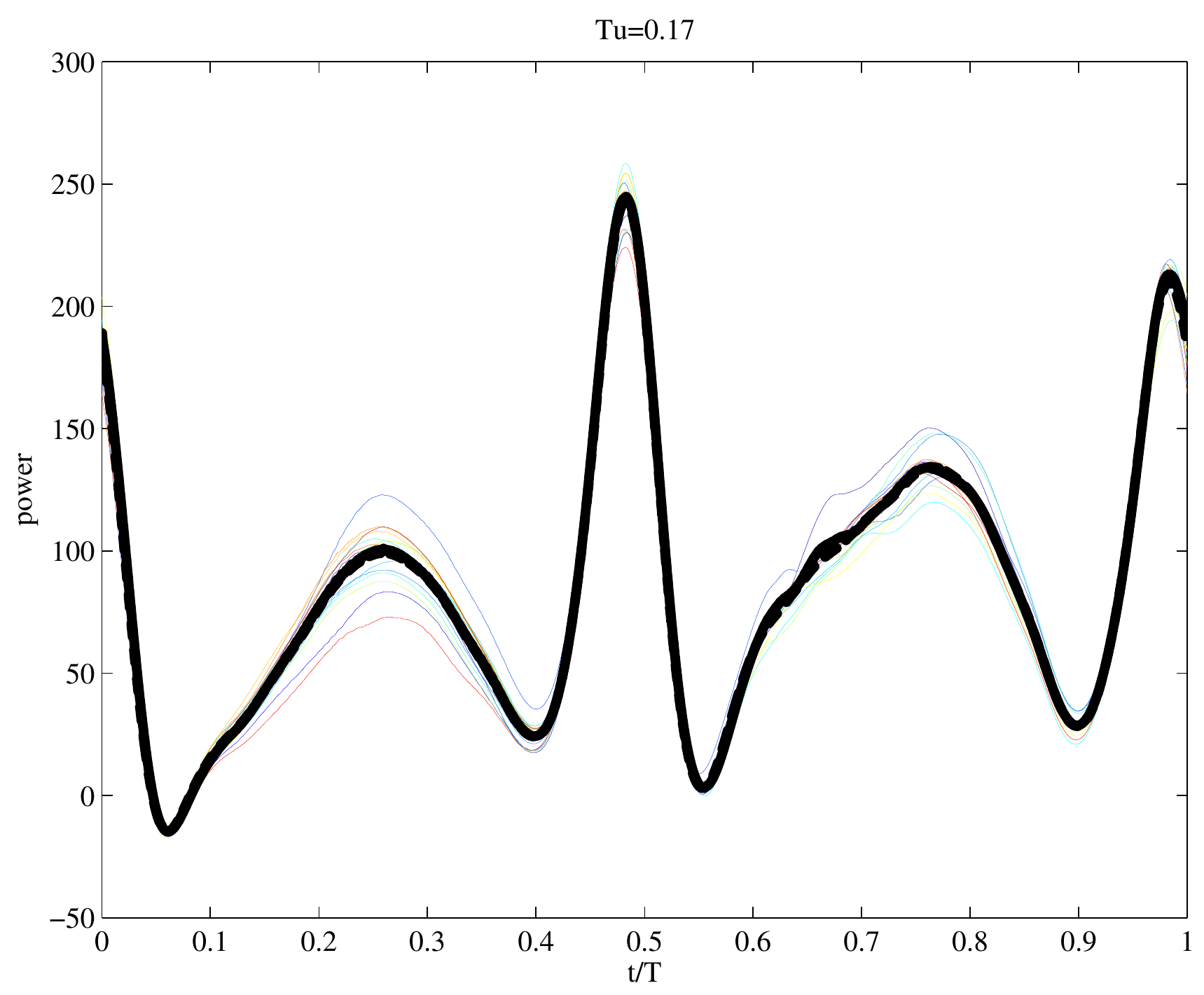}\includegraphics[width=0.5\textwidth]{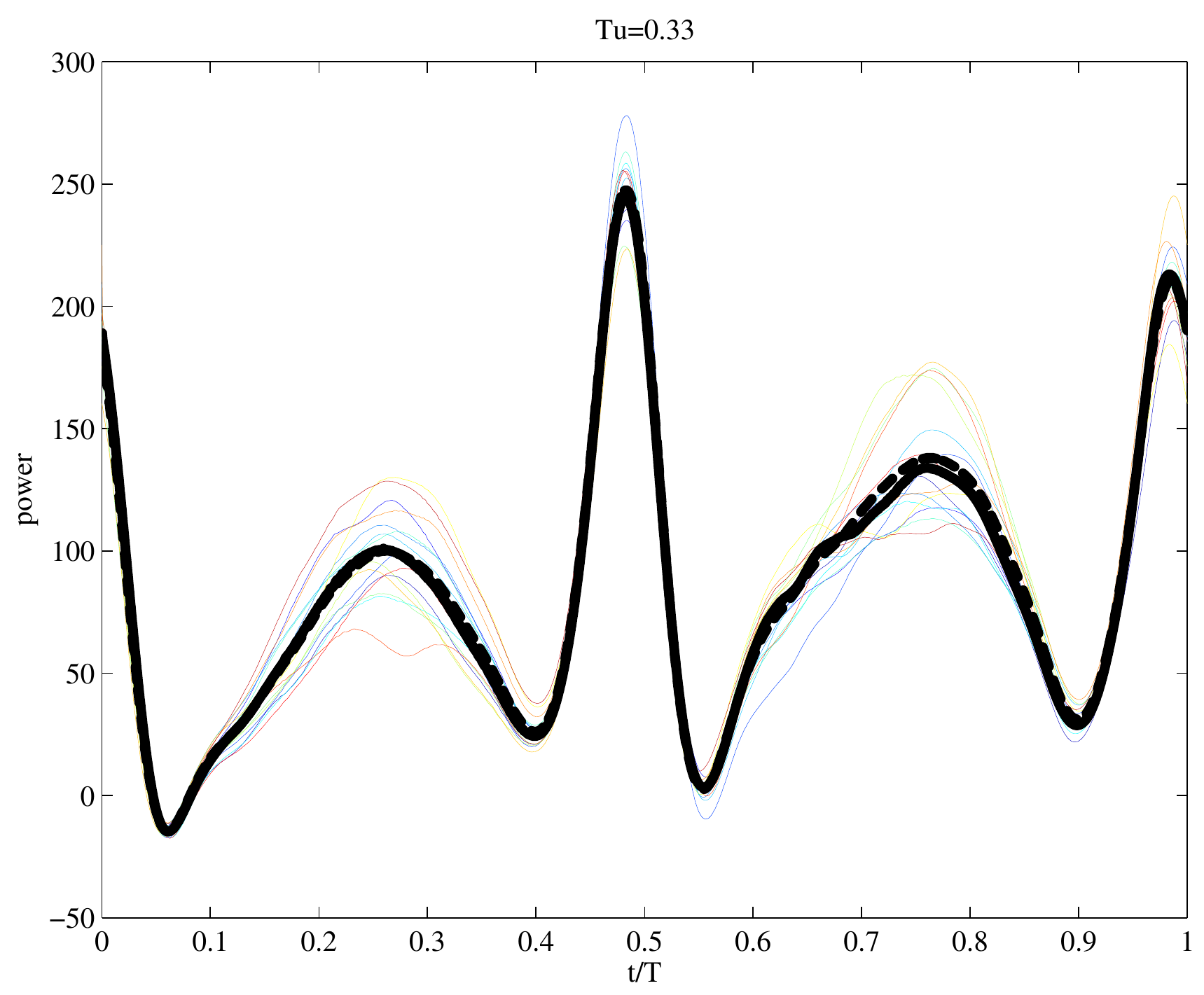}
\par\end{centering}

\begin{centering}
\includegraphics[width=0.5\textwidth]{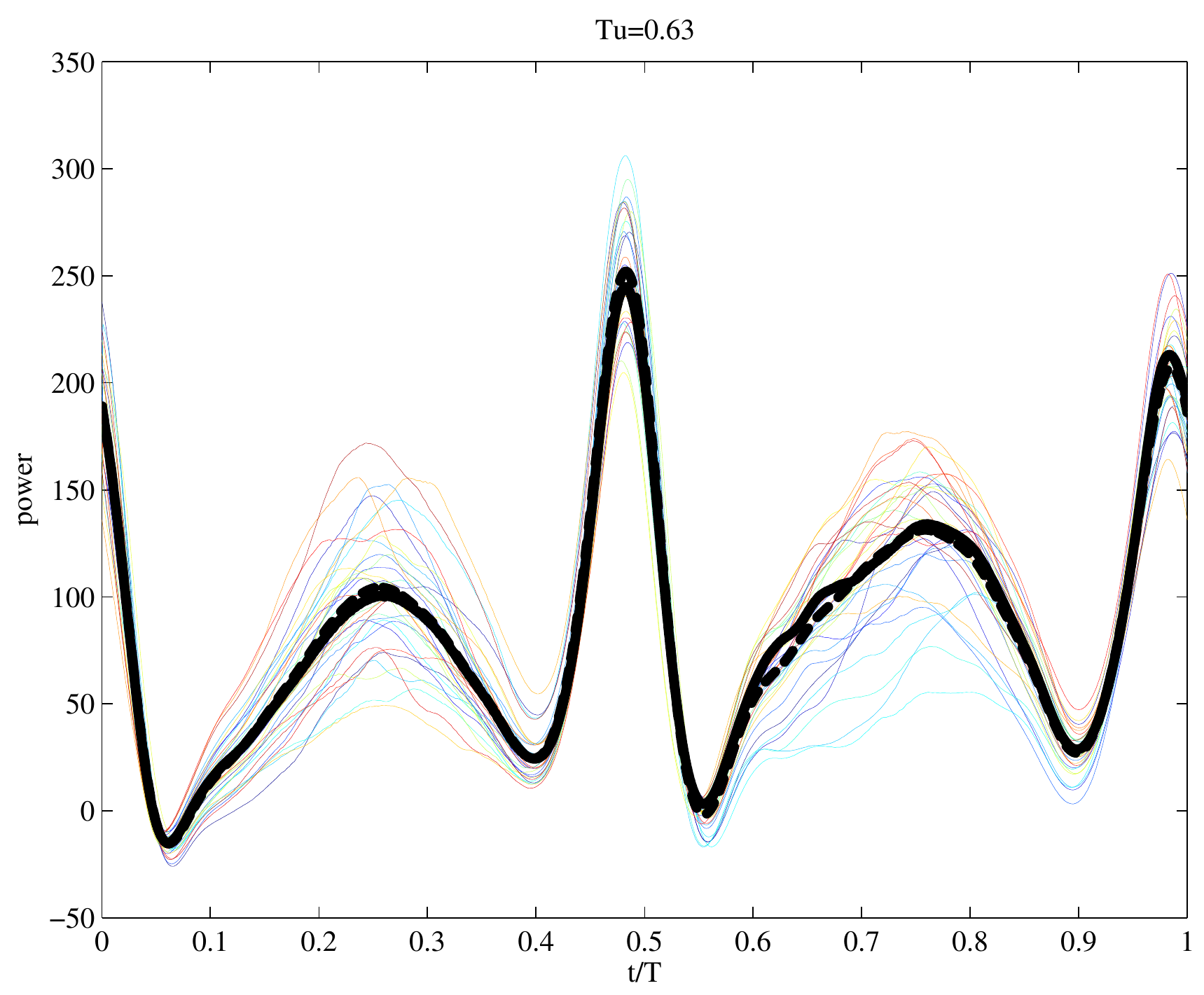}\includegraphics[width=0.5\textwidth]{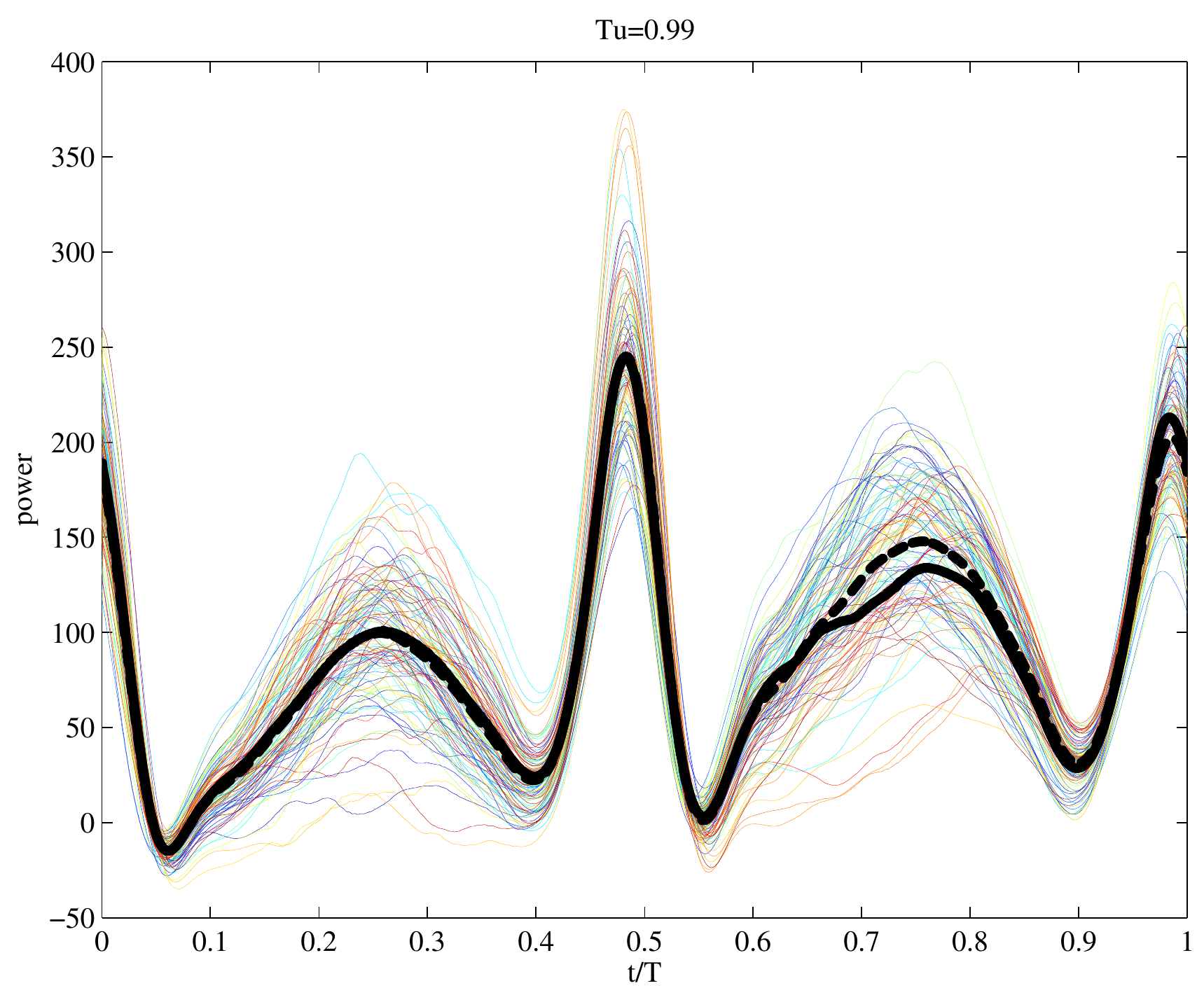}
\par\end{centering}

\caption{Aerodynamic power {[}W / kg body mass{]} as a function of wing beat
fraction time $t/T$, for different values of $Tu=\left\{ 0.17,0.33,0.63,0.99\right\} $.
Thin colored lines represent individual realizations for the given
turbulence intensity. The average over all turbulent realizations
for the given $Tu$ is represented by the thick black dashed line
and the laminar case by the thick black continuous line. The averaged
turbulent and laminar curves virtually coincide.}
\end{figure}

\begin{figure}
\begin{centering}
\includegraphics[width=0.5\textwidth]{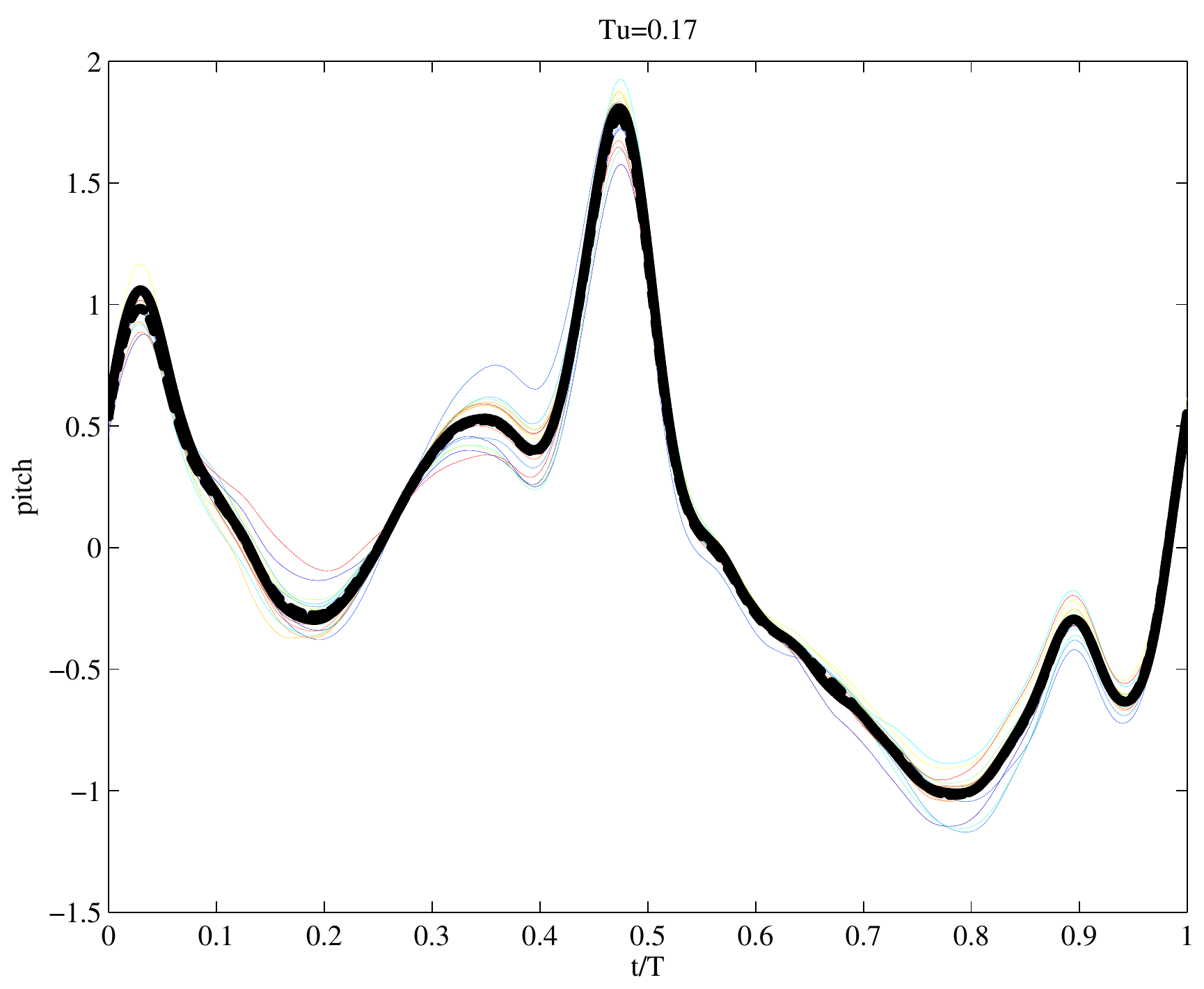}\includegraphics[width=0.5\textwidth]{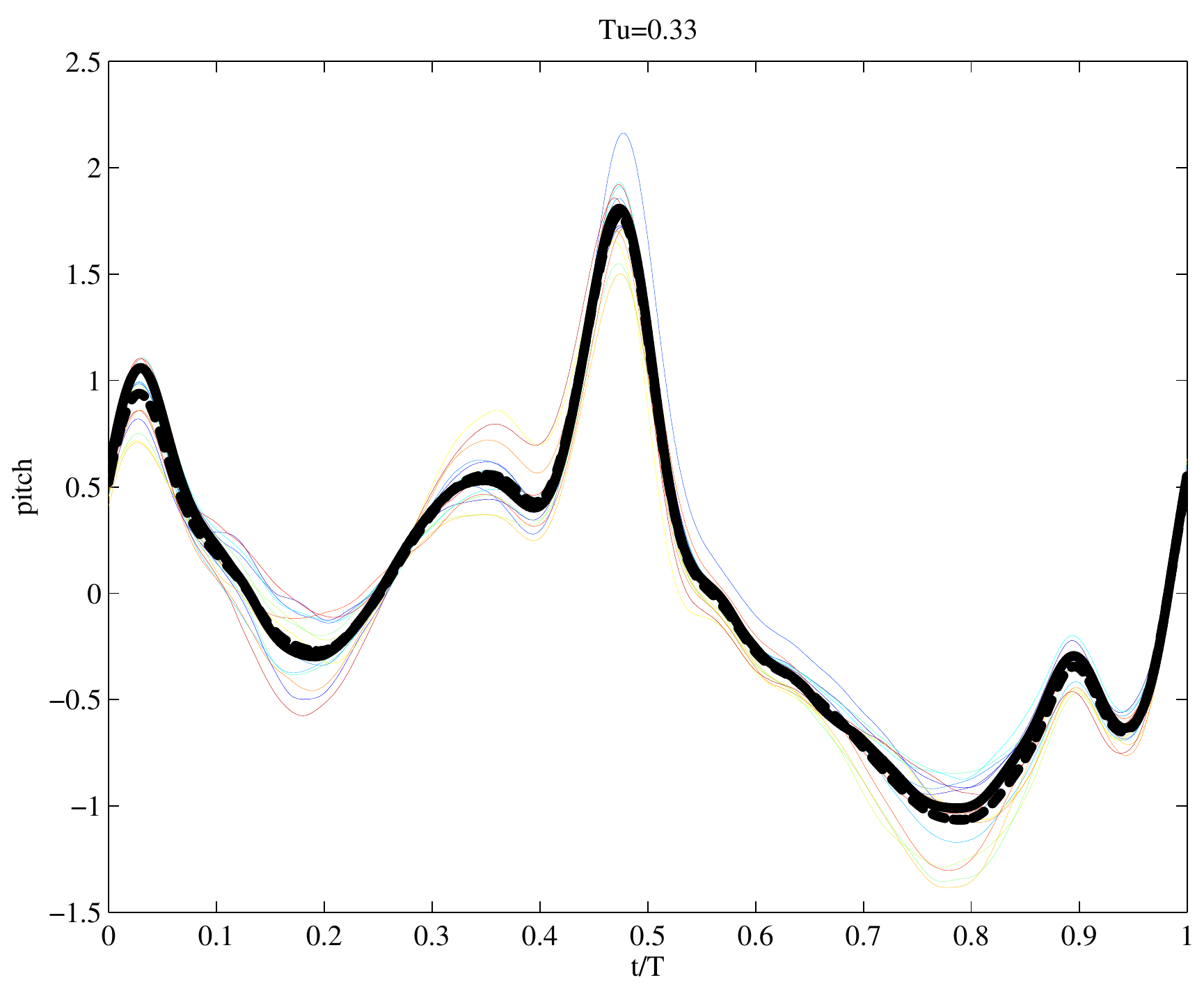}
\par\end{centering}

\begin{centering}
\includegraphics[width=0.5\textwidth]{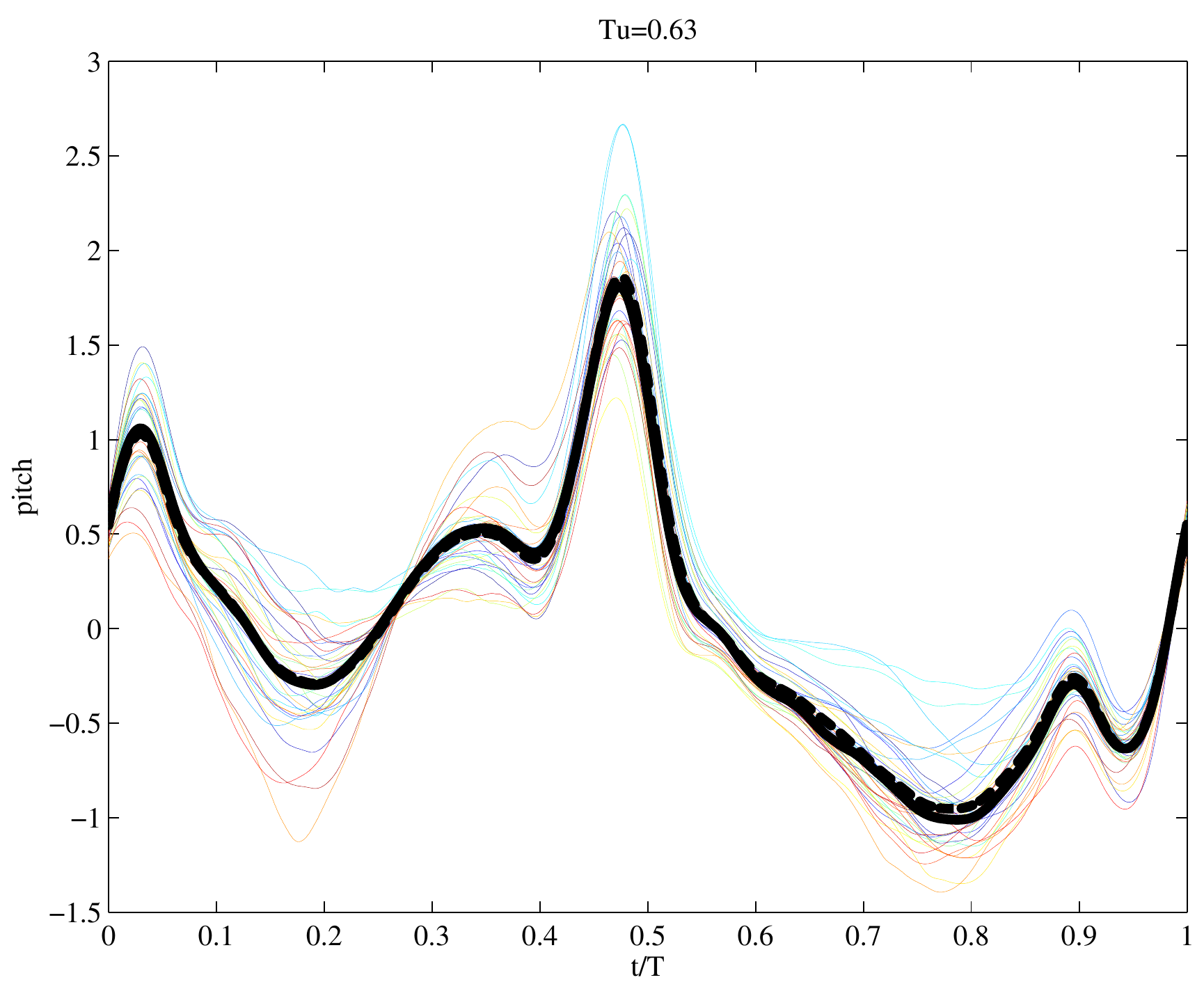}\includegraphics[width=0.5\textwidth]{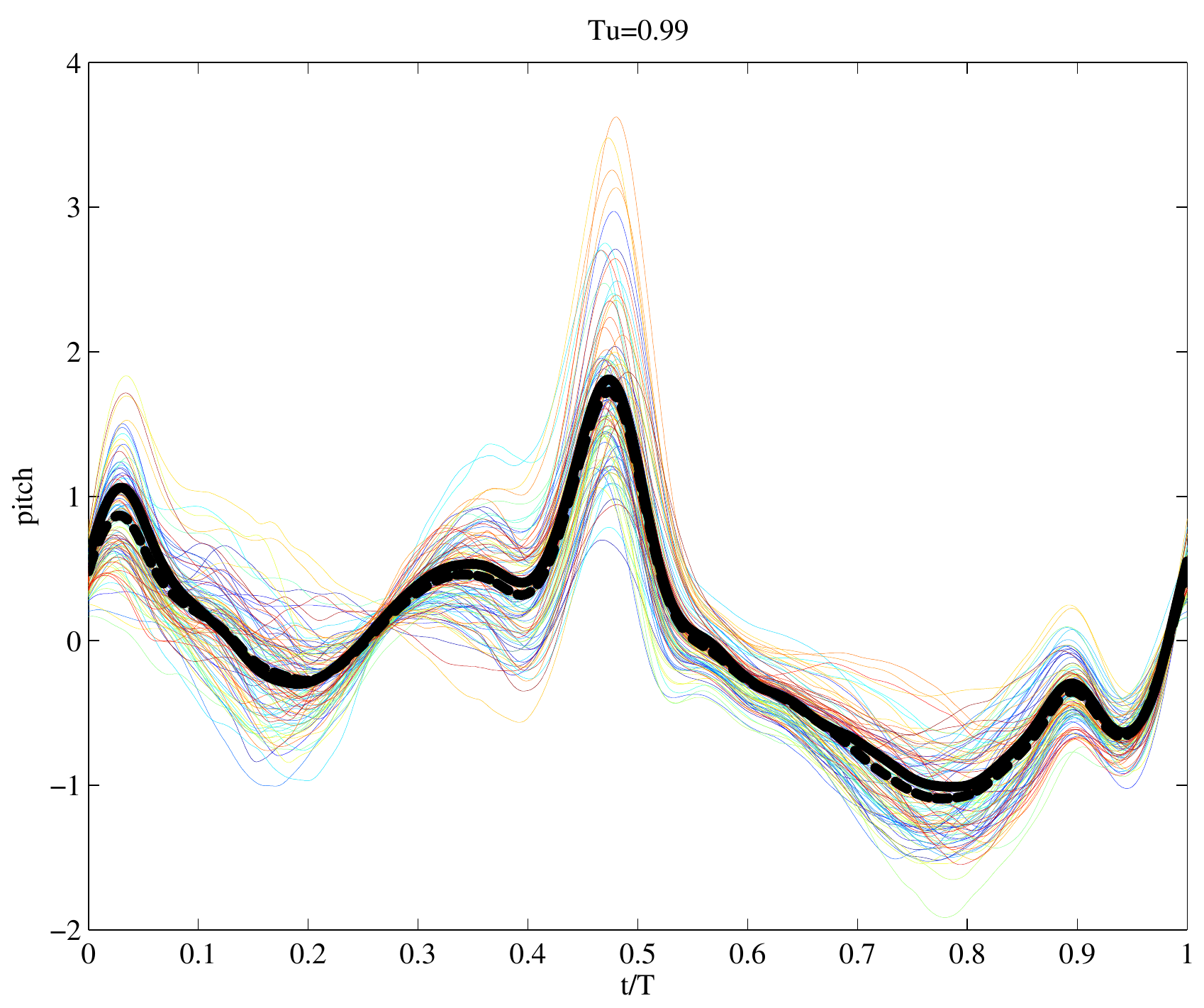}
\par\end{centering}

\caption{Pitch moment (normalized by weight force times wing length) as a function
of wing beat fraction time $t/T$, for different values of $Tu=\left\{ 0.17,0.33,0.63,0.99\right\} $.
Thin colored lines represent individual realizations for the given
turbulence intensity. The average over all turbulent realizations
for the given $Tu$ is represented by the thick black dashed line
and the laminar case by the thick black continuous line. The averaged
turbulent and laminar curves virtually coincide.}
\end{figure}

\subsection{Generalization of results: Supporting results for a different insect
model}

The results presented in the main text were obtained using the model
bumblebee described below. However the results are expected to be
generalizable to a wider range of flapping insects in a similar Reynolds
number regime. The typical leading edge vortex has been observed even
in bird flight, however our results are likely not applicable for
very large flyers that might partially glide. For the smallest insects,
the effect of ambient turbulence may be reduced to a quasi-steady
large-scale perturbation.

\begin{figure}
\begin{centering}
\includegraphics[scale=0.75]{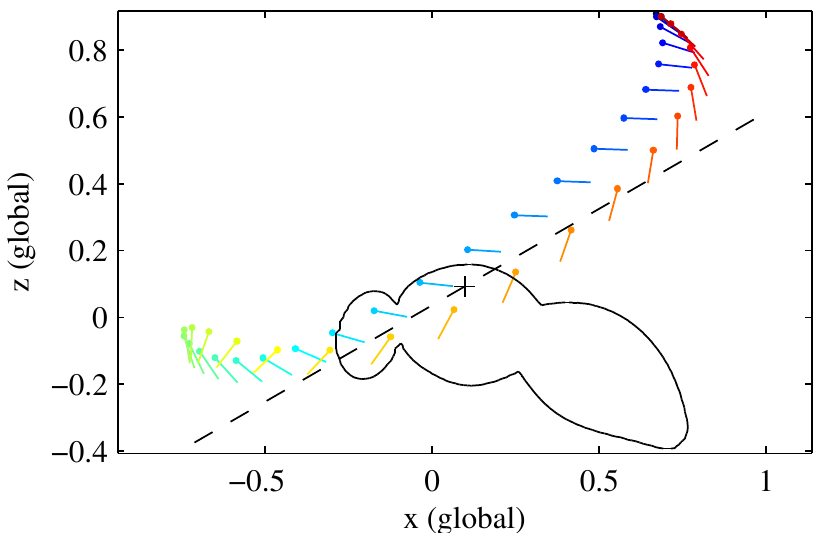}\includegraphics[scale=0.75]{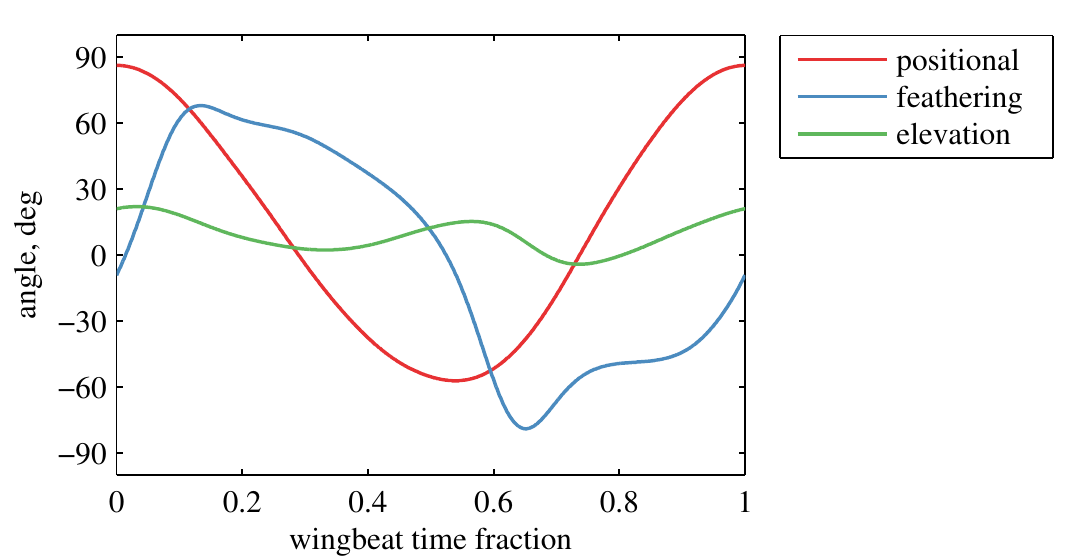}
\par\end{centering}

\caption{\label{fig:Model-description-for}Morphology for the model housefly.
Left: visualization of the wingbeat. Right: wingbeat kinematics as
a function of $t/T$.}

\end{figure}

To further support our results, we consider an additional insect model.
It geometry is based on a fruitfly \cite{Maeda2013}, the wingbeat
kinematics is adapted from \cite{Fry2005}, where hovering flight
was considered. The positional, feathering and elevation angle of
the wing is visualized in figure \ref{fig:Model-description-for}
(right). Since \cite{Fry2005} considered hovering flight, we increased
the body pitch angle by $\beta=-15^{\circ}$ and the stroke plane
angle by $\eta=-45^{\circ}$, as visualized in figure \ref{fig:Model-description-for}
(left). The body is thus more horizontal and the stroke plane is inclined
with respect to horizontal. The Reynolds number is set to $Re=1296$,
which is thus $10$ times as high as for a fruitfly. We therefore
refer to this model as a model housefly.

The model housefly is, unlike the model bumbblebee, not designed to
mimic a real animal. Its purpose is rather to illustrate that the
stability of the leading edge vortex in turbulent inflow is a generic
feature, and not limited to the bumblebee. For this reason, we give
all quantities in this subsection in dimensionless units, using $R,f$
and $\varrho_{f}$ as normalization. 

Except the insect model, the free-stream velocity $u_{\infty}=1.2$
and the viscosity $\nu=1.16\cdot10^{-3}$, the setup is unaltered.
The same resolution and time stepping is used, and likewise a series
of runs with different turbulent inflow fields with the same statistical
properties is performed. We fix the turbulence intensity to $Tu=0.44$
(with $\Lambda=0.771$, $\lambda=0.220$, $\ell_{\eta}=0.011$, $Re_{\lambda}=100$,
$N_{R}=9$, $N_{w}=36$).

\begin{figure}
\begin{centering}
\includegraphics[width=0.6\textwidth]{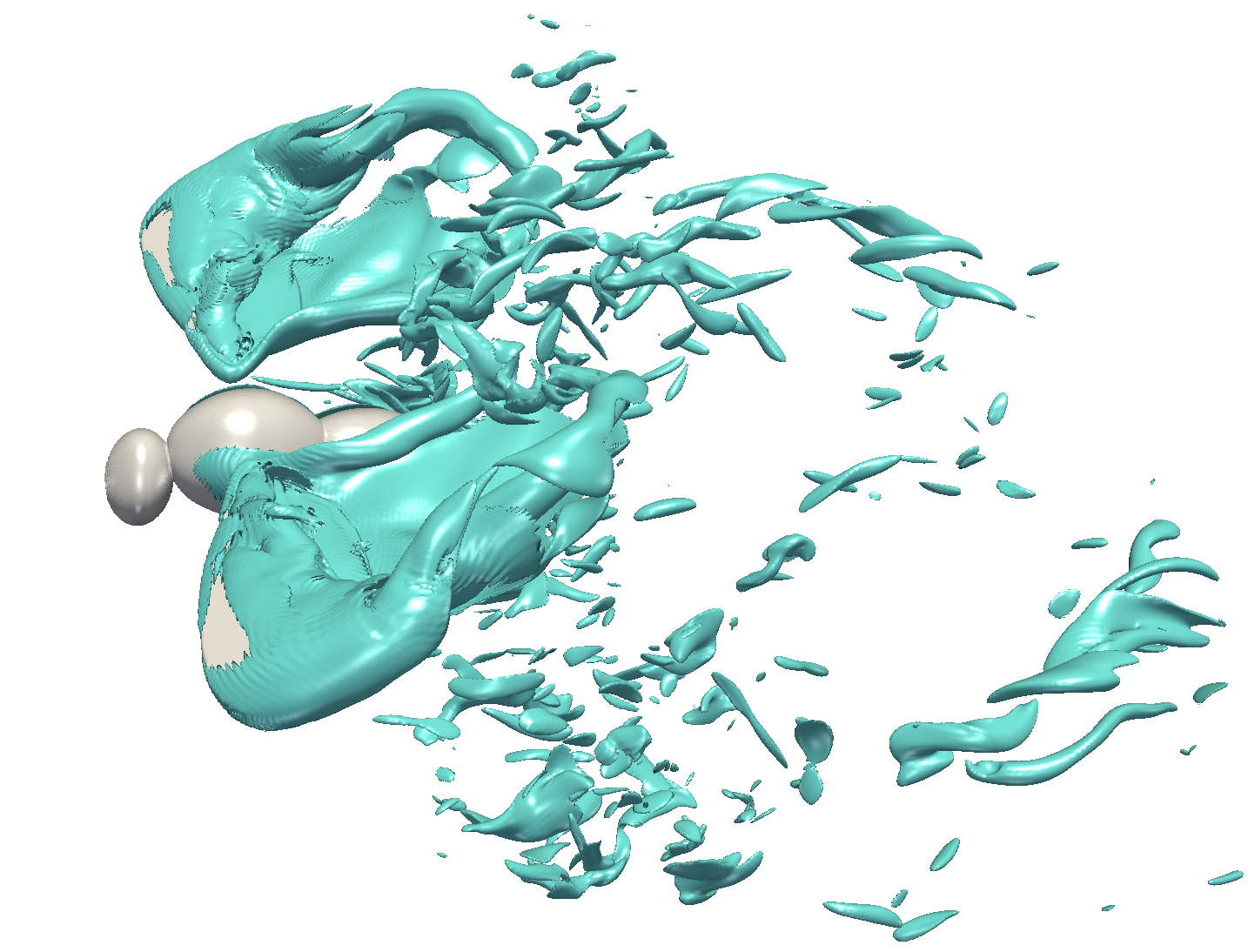}
\par\end{centering}

\caption{Flow field generated by the model housefly under laminar inflow condition.
Shown is the $\left\Vert \underline{\omega}\right\Vert =100$ isosurface
of vorticity magnitude.\label{fig:Flow-field-generated}}

\end{figure}

As for the bumblebee in the main text, we first visualize the model
in laminar inflow. Visual inspection of the $\left\Vert \underline{\omega}\right\Vert =100$
isosurface in figure \ref{fig:Flow-field-generated} reveals features
similar to the bumblebee: a pronounced leading edge vortex is visible,
connected to larger wingtip vortices. Owing to the reduced Reynolds
number, the flow field presents less small scales then in the case
of the bumblebee (see Movie \ref{fig:Wake-generated-by-BB}).

To limit computational efforts, we study only one turbulence intensity
for this model. Table \ref{tab:Data-from-the-experiments} summarizes
the results obtained in laminar and turbulent inflow.

\begin{table}
\begin{centering}
{\small{}}%
\begin{tabular*}{1\textwidth}{@{\extracolsep{\fill}}ccccccc}
\toprule 
\multirow{1}{*}{{\small{}$Tu$}} & \multicolumn{1}{c}{{\small{}}%
\begin{minipage}[t]{2cm}%
{\small{}Forward force $F_{h}$}%
\end{minipage}} & \multicolumn{1}{c}{{\small{}}%
\begin{minipage}[t]{2cm}%
{\small{}Vertical force $F_{v}$}%
\end{minipage}} & \multicolumn{1}{c}{{\small{}}%
\begin{minipage}[t]{2cm}%
{\small{}Aerodynamic power $P_{\mathrm{aero}}$}%
\end{minipage}} & \multicolumn{1}{c}{{\small{}}%
\begin{minipage}[t]{2cm}%
{\small{}Moment $M_{x}$ (roll)}%
\end{minipage}} & \multicolumn{1}{c}{{\small{}}%
\begin{minipage}[t]{2cm}%
{\small{}Moment $M_{y}$ (pitch)}%
\end{minipage}} & \multicolumn{1}{c}{{\small{}}%
\begin{minipage}[t]{2cm}%
{\small{}Moment $M_{z}$ (yaw)}%
\end{minipage}}\tabularnewline
\midrule
\midrule 
{\small{}$0$} & {\small{}$-0.94$} & {\small{}$4.99$} & {\small{}$19.15$} & {\small{}$0.00$} & {\small{}$-0.69$} & {\small{}$0.00$}\tabularnewline
\midrule
\midrule 
{\small{}$0.44$} & {\small{}$-1.05{}^{\pm0.09}\pm0.26$} & {\small{}$5.00{}^{\pm0.15}\pm0.46$} & {\small{}$19.18^{\pm0.31}\pm0.96$} & {\small{}$-0.01^{\pm0.05}\pm0.16$} & {\small{}$-0.70^{\pm0.04}\pm0.13$} & {\small{}$-0.04^{\pm0.07}\pm0.22$}\tabularnewline
\bottomrule
\end{tabular*}
\par\end{centering}{\small \par}

\caption{Model housefly. Aerodynamic forces, power and moments obtained in
the numerical experiments. All quantities are dimensionless (using
$R,f,\varrho_{\infty}$ as reference values). Values are given by
mean value $\overline{x}$, 95\% confidence interval $\delta_{95}$
and standard deviation $\sigma$ in the form $\overline{x}^{\pm\delta_{95}}\pm\sigma$.
Results were obtained by performing $N_{R}=9$ simulations, yielding
$N_{w}=36$ statistically independent wingbeats.\label{tab:Data-from-the-experiments}}
\end{table}

\begin{figure}
\begin{centering}
\includegraphics[width=0.5\textwidth]{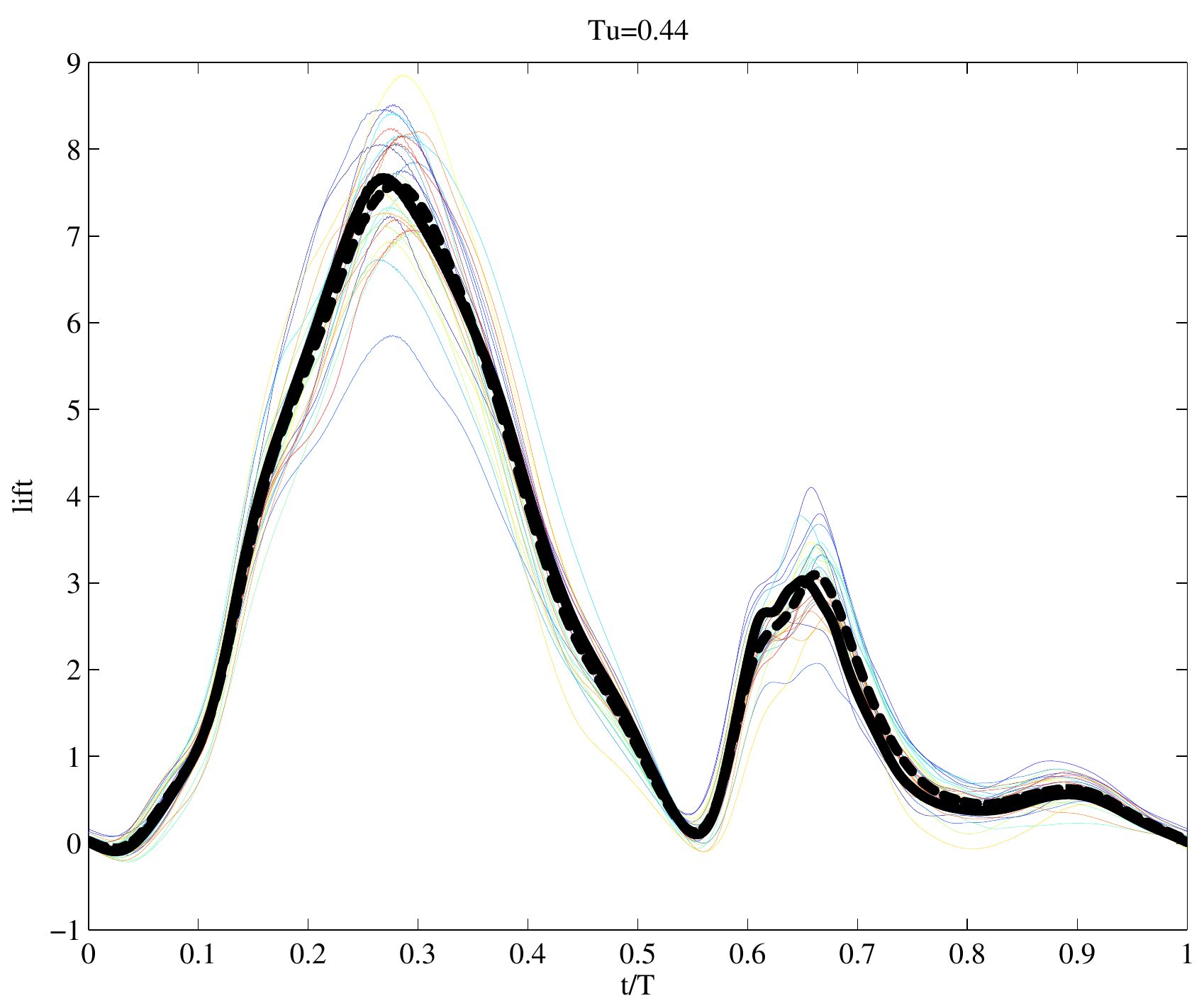}\includegraphics[width=0.5\textwidth]{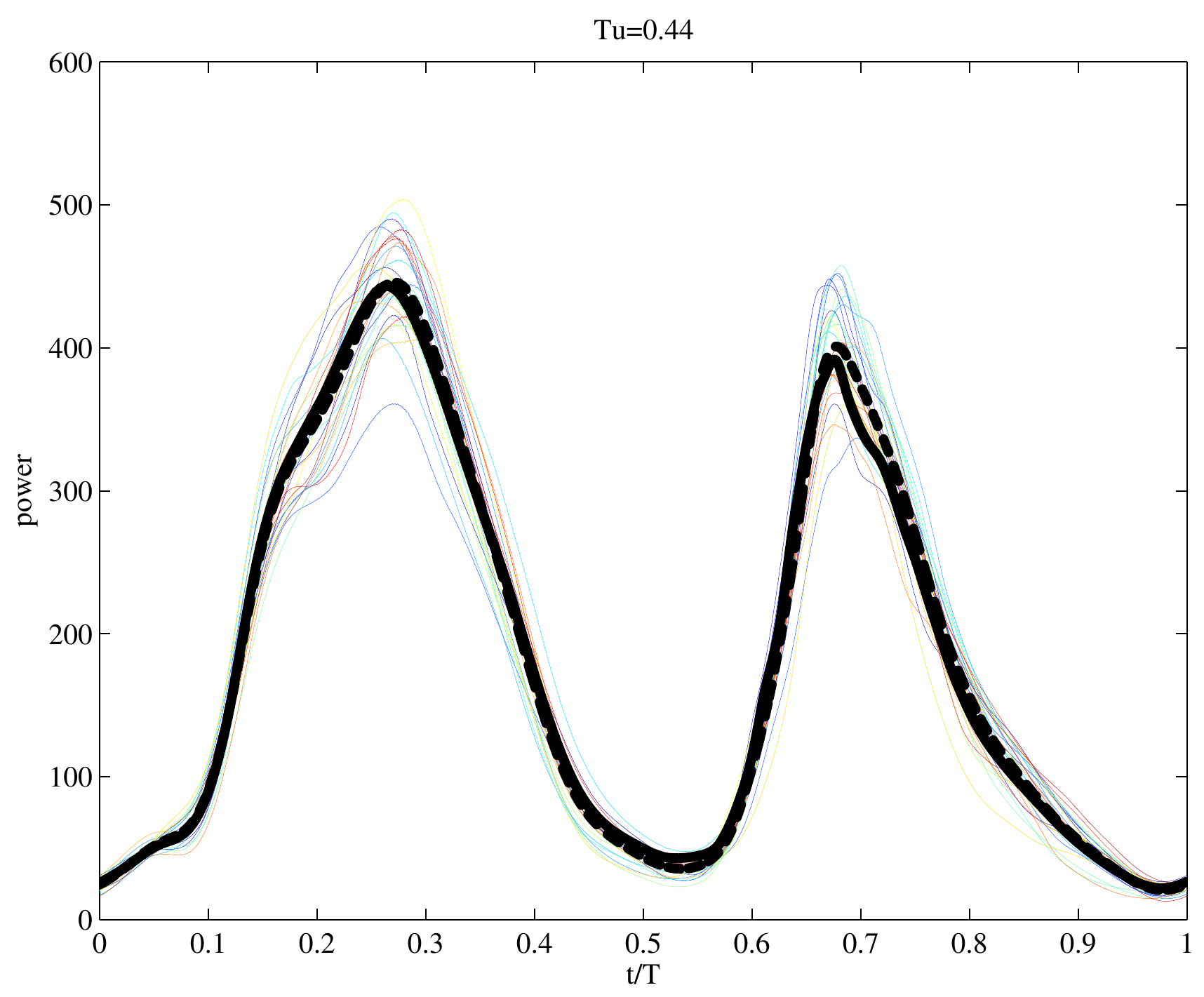}
\par\end{centering}

\caption{Model housefly. Instantaneous lift force (left) and aerodynamic power
(right), both in dimensionless units (using $R,f,\varrho_{\infty}$
as reference values). The averaged curve (thick black dashed line)
and the one of the laminar case (thick black continuous line) almost
overlap.\label{fig:Instantaneous-lift-force}}

\end{figure}

The conclusions from these simulations are similar to those drawn
from the bumblebee. The average forces are not significantly different
compared to the laminar case, and thus the underlying aerodynamic
mechanisms are robust with respect to turbulent perturbations. Finally,
we visualize in figure \ref{fig:Instantaneous-lift-force} the instantaneous
lift force and aerodynamic power. Visibily, as in the previous figures,
the ensemble-averaged time series is in good agreement with the laminar
curve.

\section{Materials And Methods}

\subsection{Model bumblebee \label{sub:Model-bumblebee}}

\begin{figure}
\begin{centering}
\includegraphics[width=0.9\textwidth]{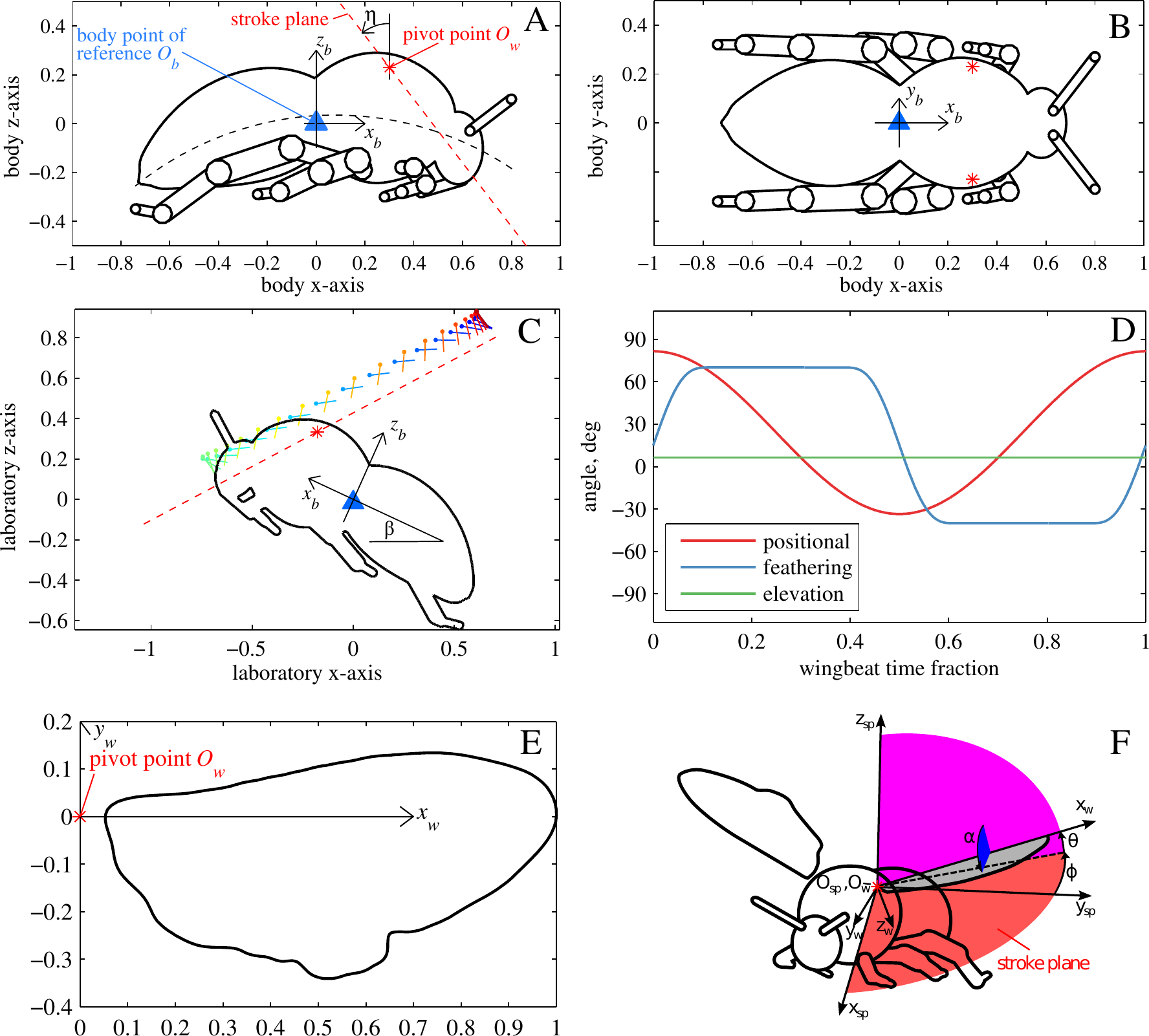}
\par\end{centering}

\caption{Bumblebee model. (A-B) side and top view of the bumblebee body with
the definitions of the body point of reference and the wing pivot
points. The body is obtained by sweeping an elliptical section along
an arc (black dashed line). The wings rotate around the pivot points
(red stars) within the stroke plane, which is inclined by $\eta=37.5^{\circ}$
with respect to the body normal. The entire body is inclined by $\beta=24.5^{\circ}$
with respect to the horizontal plane (C). The two wings follow the
symmetric motion protocol illustrated in D. The wing geometry is shown
in E, where the distance pivot point - wing tip is normalized to unity.
Wing angles are defined in F. \label{fig:kinematic-bumblebee-model}}
\end{figure}

The parameters of the bumblebee model were derived from \cite{Dudley1990},
case BB01. The animal's body mass, $m$, is approximately 175 mg and
the wing length, $R$, amounts to 13.2 mm. We assumed the insect to
be composed of linked rigid bodies and modeled the insect's body shape
by sweeping an elliptical section of variable size along a curvilinear
centerline. Compared to other insects, bumblebees have relatively
thick legs that potentially create non-negligible aerodynamic effects.
We thus included all legs, the proboscis and the antennae in our model
as circular cylindrical sections joined by spheres, and further assumed
bilateral symmetry of the insect. Fig. \ref{fig:kinematic-bumblebee-model}A-B
show side and top views of the modeled animal with all linear dimensions
normalized to wing length. We numerically computed the inertia tensor,
$\underset{=}{I}$, of the bumblebee neglecting the wings and assuming
uniform body density of $\varrho_{b}=m/V=362\,\text{kg}/\text{m}^{3}$,
where $V=0.48\,\text{c}\text{m}^{3}$, from equation 
\begin{eqnarray*}
\underset{=}{I} & = & \varrho_{b}\int_{V}\left(\left(\underline{r}\cdot\underline{r}\right)\underset{=}{E}-\underline{r}\otimes\underline{r}\right)dV\\
 & = & \left(\begin{array}{ccc}
0.183 & 0 & 0.1307\\
0 & 0.400 & 0\\
0.1307 & 0 & 0.339
\end{array}\right).
\end{eqnarray*}
Diagonalizing the inertia tensor yields the principal moments of inertia
$I_{11}=0.1092$, $I_{22}=0.3998$ and $I_{33}=0.4136$, all in units
of $10^{-8}\,\text{kg}\text{m}^{2}$. The data show that moment of
inertia about the roll axis ($x$-axis) is approximately four times
smaller than about the yaw and pitch axes, which is in agreement with
\cite{Ravi2013} and \cite{Combes2009}. 

The wing contour was digitized from \cite{UMIC} and the area scaled
to $48.37\,\text{m}\text{m}^{2}$ of a single wing and $c=3.66$ mm
mean wing chord (Fig. \ref{fig:kinematic-bumblebee-model}E). Following
\cite{Xu2013}, we modeled wing kinematics from sinusoidal changes
in position angle $\phi\left(t/T\right)=\overline{\phi}+\Phi\sin\left(2\pi t/T\right)$
with mean $\overline{\phi}=24^{\circ}$ and an amplitude $\Phi=115^{\circ}$,
using a constant elevation angle $\theta=12.55^{\circ}$ with respect
to the stroke plane, and assuming a constant feathering angle $\alpha$
of $70^{\circ}$ during the upstroke and $-40^{\circ}$ during the
downstroke. At the ventral and dorsal stroke reversal, $\alpha$ changed
sinusoidally over a duration of 0.22 cycle durations. The wingbeat
kinematics are visualized in Fig. \ref{fig:kinematic-bumblebee-model}D.
The inclination angle, $\eta$, of the stroke plane against the longitudinal
body axis is $37.5^{\circ}$, and the body pitch angle, $\beta$ is
$24.5^{\circ}$ with respect to the horizontal plane (Fig. \ref{fig:kinematic-bumblebee-model}C).
We modeled the simplified kinematics at an intermediate flight speed
of 2.5m/s and a wingbeat frequency $f$ of 152 Hz according to values
previously measured in freely flying bumblebees \cite{Xu2013,Dudley1990}.
The modeled wings flapped at mean wing tip velocity of $U=2\Phi Rf=8.75\,\text{m/s}$
and a Reynolds number of 2042 based on $U$, $c$ and the kinematic
viscosity of air at $300\,\text{K}$, $\nu=1.57\cdot10^{-5}\,\text{m}^{2}/\text{s}$.
The fluid density is $\varrho_{f}=1.177\,\text{kg}/\text{m}^{3}$.

\subsection{Numerical method}

The challenge when simulating insects in turbulence is to accurately
resolve a multitude of temporal and spatial scales while taking into
account the insect as a complex, time-dependent geometry. To this
end, the volume penalization method \cite{Angot1999} is employed.
Thus we solve the incompressible Navier-Stokes equations,
\begin{eqnarray}
\partial_{t}\underline{u}+\underline{\omega}\times\underline{u} & = & -\nabla p+\frac{1}{\mathrm{Re}}\nabla^{2}\underline{u}-\frac{\chi}{C_{\eta}}\left(\underline{u}-\underline{u}_{s}\right)\label{eq:NST}\\
\nabla\cdot\underline{u} & = & 0,\label{eq:div-free}
\end{eqnarray}
written in dimensionless form. The penalization term $-\chi/C_{\eta}\left(\underline{u}-\underline{u}_{s}\right)$
has been added to enforce the no-slip boundary condition $\underline{u}=\underline{u}_{s}$
in the solid. The geometry is encoded in the indicator function $\chi$,
which vanishes in the fluid and equals unity in the solid. The penalization
parameter is $C_{\eta}=2.5\cdot10^{-4}$. Since the geometry changes
in time, we use a smooth $\chi$-function that is resampled on the
Eulerian fluid grid in every time step \cite{Engels2014}. Equations
(\ref{eq:NST}-\ref{eq:div-free}) are solved on an equidistant uniform
Cartesian grid of $1152\times768\times768$ points, using a Fourier
pseudospectral discretization \cite{Engels2014,Kolomenskiy2011a}.
This method is particularly well adapted for turbulence simulations,
because of vanishing numerical diffusion and dispersion. The implementation
uses highly efficient computational FFT-libraries \cite{Pekurovsky2012},
suitable for massively parallel supercomputers. Results presented
here are obtained using up to 32768 computing threads on a Blue Gene-Q
machine located at IDRIS, Orsay, France. Details about the numerical
method and our code for flapping insect flight can be found in \cite{Engels2015a}.

\subsection{Inflow turbulence}

Insects successfully fly in turbulent environments \cite{Ortega-Jimenez2013,Ravi2013,Vance2013}.
Since the properties of these aerial perturbations depend on a large
number of parameters, we model them by homogeneous isotropic turbulence
(HIT). This is a reasonable assumption for the small turbulent scales
relevant to insects. The turbulence fields, $\widetilde{\underline{u}}'$,
were pre-computed in a periodic computational domain of size $\left(2\pi\right)^{3}$,
and we denote quantities from the HIT simulation with the tilde overset.
In such a turbulence simulation, we initialized the velocity field
at time $\widetilde{t}=0$ as a random velocity field with given energy
spectrum \cite{Rogallo1981}. This velocity field evolved according
to the incompressible Navier-Stokes equations, with the energy dissipation
compensated by a forcing term \cite{Kaneda2003} that continuously
fed energy into the largest fluid scales, i.e., the lowest wavenumbers.
After reaching the statistically steady state, independent of the
initial condition, we periodically saved snapshots of the velocity
field $\widetilde{\underline{u}}'\left(x,y,z\right)$.

Different turbulence intensities were computed by modifying the turbulent
Reynolds number $R_{\lambda}=\widetilde{U}\widetilde{\lambda}/\widetilde{\nu}$,
where $\widetilde{\lambda}$ is the Taylor-micro scale, defined as
$\widetilde{\lambda}=\left(15\widetilde{\nu}\widetilde{U}^{2}/\widetilde{\varepsilon}\right)^{1/2}$,
$\widetilde{U}$ the RMS velocity, $\widetilde{\varepsilon}$ the
dissipation rate, and $\widetilde{\nu}$ the dimensionless kinematic
viscosity in the HIT simulation. For each value of $R_{\lambda}$,
one computation was performed and several snapshots were stored. To
ensure that the snapshots are uncorrelated, the saving interval was
set to $10\widetilde{T}_{0}$, where $\widetilde{T}_{0}$ is the eddy
turnover time, $\widetilde{\Lambda}/\widetilde{U}$, and $\widetilde{\Lambda}$
is the integral length scale, $\widetilde{\Lambda}=\frac{\pi}{2\widetilde{U}^{2}}\int_{0}^{k_{\mathrm{max}}}k^{-1}\widetilde{E}\left(k\right)\mathrm{d}k$.
This way we generated 44 different turbulence flow fields with parameters
assembled in table II of the main article. 

The turbulence fields were rescaled to match the normalization of
the insect simulations using the relation $\underline{u}'=\left(2\pi\nu\right)/\left(\ell_{y}\widetilde{\nu}\right)\,\widetilde{\underline{u}}'$
and then added to the imposed mean flow $u_{\infty}$ in the inlet
region of the computational domain (Fig. \ref{fig:setup}). The scaling
relation has one degree of freedom, which is the lateral size of the
domain, $\ell_{y}$. Varying $\ell_{y}$ is equivalent to changing
the animal's size relative to the length scales of the turbulence
field. Thus the parameter $\ell_{y}$ needs to be large enough to
reduce the effect of periodicity in the lateral direction and small
enough to produce turbulent length scales similar to natural perturbations.
We thus used an intermediate value for the lateral size of $4R$.

\subsection{Numerical wind tunnel}

\begin{figure}
\begin{centering}
\includegraphics{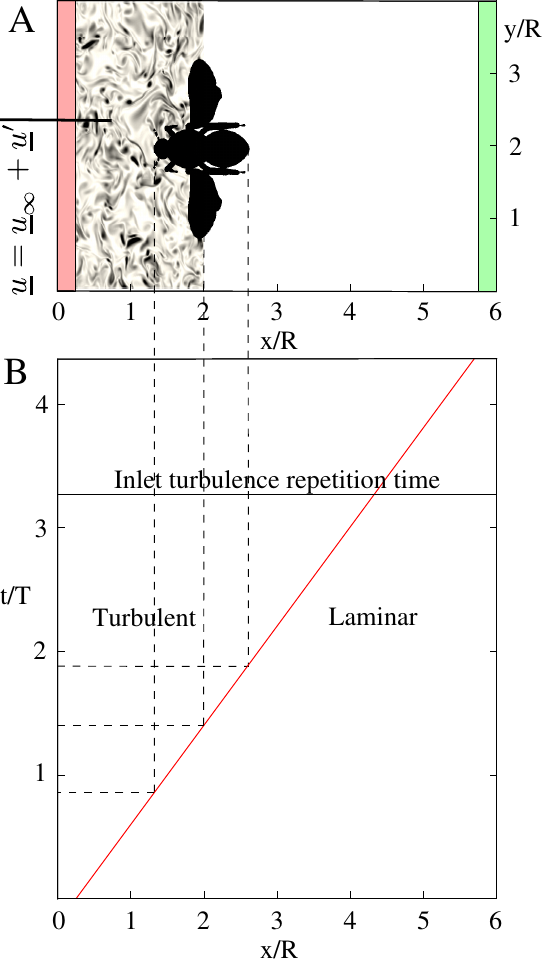}
\par\end{centering}

\caption{\label{fig:setup}Sketch of the computational setup with the insect
tethered at $x/R=2$. The red shaded area indicates where the inflow,
consisting of a uniform mean flow and turbulent fluctuations, $\underline{u}=\underline{u}_{\infty}+\underline{u}'$,
is imposed. The green area is the outflow vorticity sponge \cite{Engels2014}.
(B) Propagation of the turbulent front (red line) in the space-time
diagram. The inflow turbulence $\underline{u}'$ field repeats after
$t/T=3.21$.}
\end{figure}

The design of the simulated wind tunnel is illustrated in Fig. \ref{fig:setup}A-B.
At the outlet, a vorticity sponge with a thickness of 48 grid points
gradually absorbs the wake to model the appropriate outflow condition
\cite{Engels2014}. At the inlet, isotropic turbulent fluctuations
are added in a slice of 48 grid points thickness to a uniform freestream
of 2.5 m/s. The pre-computed turbulence field is imposed by performing
a change of variables, $z=u_{\infty}t/T$. Since the HIT computation
is periodic, the injected flow field repeats after $t_{rep}=\ell_{y}/u_{\infty}=3.21T$.
The initial condition corresponds to unperturbed laminar flow, thus
the turbulent--laminar interface travels downstream, as illustrated
in Fig. \ref{fig:setup}A, and turbulence hits the head and tail of
the model bumblebee approximately after 0.95 and 2 wing strokes, respectively.
Throughout the remaining computational sequence, the insect is entirely
immersed in the turbulent flow. Owing to laminar flow conditions,
we thus excluded the first two wing strokes of each simulation from
the statistical analysis. 

Each simulation run was repeated $N_{R}$ times (table II of main
article) and yielded four subsequent, uncorrelated wing strokes. The
simulation stopped after six wing strokes because the following strokes
are no longer independent owing to the repetition of the imposed inflow
perturbation. This procedure guaranteed confident statistical means
and variances of aerodynamic forces and power. It also ensured that
the results were not biased by a randomly present coherent structure
in the turbulent inflow field.

\subsection{Wake turbulence in laminar inflow}

Figure \ref{fig:laminar_tu} and Movie \ref{fig:Wake-generated-by-BB}
illustrate the wake generated by the bumblebee without upstream perturbations.

\begin{figure}
\begin{centering}
\includegraphics[width=1\textwidth]{laminar_tu}
\par\end{centering}

\caption{\label{fig:laminar_tu} Wake generated by the bumblebee under laminar
inflow conditions, illustrated by the $\left\{ 0.2,0.4,0.6\right\} $
isosurfaces of $Tu=\sqrt{2\overline{u'^{2}}/3}/u_{\infty}$. The mean
turbulent kinetic energy is computed as $\overline{u'^{2}}=\left(\overline{u^{2}}-\overline{u}^{2}\right)$.}
\end{figure}

\begin{figure}
\begin{centering}
\includegraphics[width=0.75\textwidth]{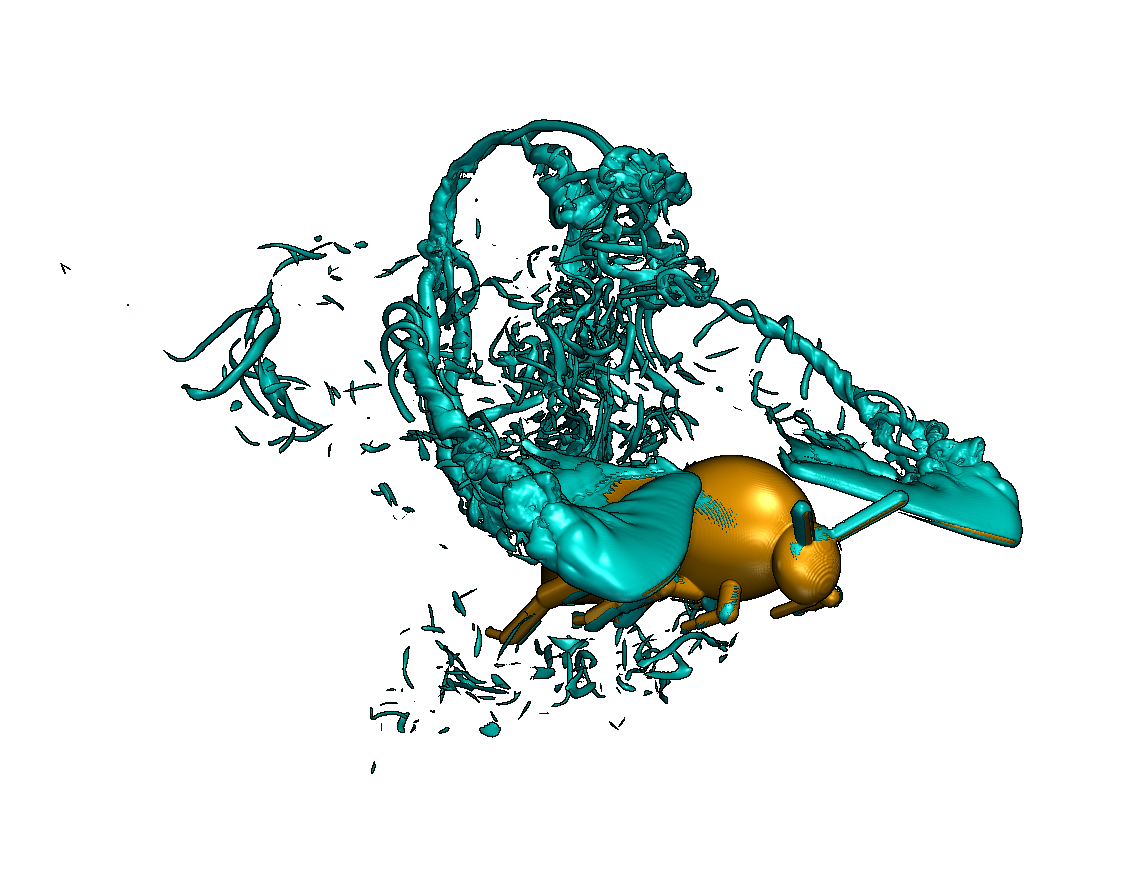}
\par\end{centering}

\setcounter{figure}{0}\renewcommand{\thefigure}{S\arabic{figure}}
\renewcommand{\figurename}{Movie }\caption{Wake generated by the bumblebee under laminar inflow conditions, illustrated
by the $\left\Vert \underline{\omega}\right\Vert =100$ isosurface
of vorticity magnitude.\label{fig:Wake-generated-by-BB}}

\renewcommand{\figurename}{figure}
\end{figure}

\bibliographystyle{apsrev4-1}
\bibliography{refs}